\documentclass[preprint, 12pt]{elsarticle}
\usepackage[a4paper,
bindingoffset=0.2in,
left=1in,
right=1in,
top=0.75in,
bottom=1in,
footskip=.25in]{geometry}

\usepackage{url}
\usepackage{fonttable}
\usepackage{ulem}
\usepackage{etoolbox}{}
\usepackage{braket}

\usepackage{amssymb}
\usepackage{amsmath}
\usepackage{url}
\usepackage{graphicx}
\usepackage{fonttable}
\usepackage{rotating}
\usepackage{ulem}
\usepackage{etoolbox}{}
\usepackage{braket}
\usepackage{epstopdf}
\usepackage{caption}
\usepackage{multirow}
\usepackage{color}

\def\be{\begin{equation}}
\def\ee{\end{equation}}
\def\bea{\begin{eqnarray}}
\def\eea{\end{eqnarray}}
\def\bdm{\begin{displaymath}}
\def\edm{\end{displaymath}}

\journal{Universe}

\begin{document}
	
	\begin{frontmatter}
		
		\title{Re-examination of the Effect of Pairing Gaps on Gamow-Teller Strength Distributions and $\beta$-decay Rates}

		\author {Jameel-Un Nabi$^{1}$}
\author{Muhammad Riaz$^{2}$}
		\author {Arslan Mehmood$^{1}$}

		\address{$^1$University of Wah, Quaid Avenue, Wah Cantt 47040, Punjab, Pakistan}
		\address{$^2$ Department of Physics, University of Jhang, Jhang, Punjab Pakistan}
		
\begin{abstract}
			
$\beta$-decay is one of the  key factors to understand the $r$-process and evolution of massive stars. The Gamow-Teller (GT) transitions drive the $\beta$-decay process. We employ the proton-neutron quasiparticle random phase approximation (pn-QRPA) model  to calculate terrestrial and stellar $\beta$-decay rates for 50 top-ranked nuclei possessing astrophysical significance according to a recent survey. The model parameters of the pn-QRPA model  affect the predicted results of $\beta$-decay. The current study investigates the effect of nucleon-nucleon pairing gaps on charge-changing transitions and the associated $\beta$ decay rates.  Three different values of pairing gaps, namely TF, 3TF and 5TF, were used in our investigation. It was concluded that both GT strength distributions and half-lives are sensitive to  pairing gap values. The 3TF pairing gap scheme, in our chosen nuclear model, resulted in best prediction with around 80$\%$ of the calculated half-lives within a factor 10 of the measured ones. The 3TF pairing scheme also led to calculation of biggest $\beta$-decay rates in stellar matter.
\end{abstract}
\begin{keyword}
Gamow-Teller strength; pairing gaps; half-lives; deformed pn-QRPA; $\beta$ decay rates, partial half lives.
\end{keyword}
		
\end{frontmatter}
	
\section{Introduction}
\label{intro} The nuclear reactions mediated by weak interactionsplay a crucial role in presupernova evolution of massive stars \cite{Bet90}. The $\beta^{\pm}$ decay, electron and positron capture are the fundamental weak interaction processes that occur during the presupernova phases. The $\beta$ decay and electron capture are transformations which produce (anti)neutrinos. A change of lepton-to-baryon fraction ($Y_e$) of the core matter affects the dynamics of collapse and subsequent explosion of the massive stars \cite{Lan03,Jan07}. Two important parameters to determine the dynamics of core-collapse are time rate of $Y_e$ and entropy of the core material \cite{Bet79}. The weak interaction mediated rates play an important role in stellar processes including hydrostatic burning and pre-supernova evolution of massive stars. The study of stellar weak interaction rates is a key area for investigation due to its significant contribution in understanding of pre-supernova evolution of massive stars. The core-collapse simulation  depends on reliable computation of ground- and excited-states Gamow-Teller (GT) strength functions \cite{Bet79}. A substantial number of unstable nuclei are present in the core with varying abundances. Weak interactions of these nuclei in stellar matter may contribute to a better understanding of the complex dynamics of core-collapse. Once an iron core develops in a giant star's later stages of evolution, there is no more fuel available to start a new burning cycle. Lepton capture and photo-disintegration processes lead to  core's increasing instability and eventual collapse. The number of electrons available for pressure support is reduced by electron capture process, whereas degeneracy pressure is enhanced during $\beta$ decays \cite{Hix03}. Few recent papers highlighting impact of $\beta$-decays on late stellar evolution include Refs.~\cite{Heg01,Mar14,Suz16,Kir19,Liu20}. 

Determination of $\beta$-decay rates is also required for the  nucleosynthesis ($s$-, $p$-, and $r$-) processes \cite{Niu22,Bur57}. The $r$-process synthesizes half of the elements heavier than iron \cite{Bur57}. The site of $r$-process remains uncertain to-date~\cite{Sma17,Wat19,Sie19}. Pre-requisites include high neutron densities and core temperatures.  In the recent years, much experimental work has been conducted to study the nuclear properties of exotic nuclei. Since the majority of these nuclei cannot be created under lab conditions, microscopic calculations of stellar weak-decay properties have gained importance in our quest to comprehend stellar processes. Numerous computations have focused on the mechanisms underlying stellar development and nucleosynthesis (e.g.,\cite{Bor06,Pan16,Tak73,Nom20,Eng99,Kum20}). 

The $\beta$-decay half-lives were estimated with the help of gross theory~\cite{Tak73}. With the advancement of computing and new technologies, calculation of ground and excited states GT strength distributions gained attention of many researchers. The charge-changing reaction rates in stellar environment were estimated using several nuclear models. Fuller, Fowler, and Newman made the first substantial effort to compute the astrophysical rates using the independent particle model (IPM) \cite{Ful80}.  To enhance the reliability of their calculation, they took into account the measurable data that was available at the time. Later many other sophisticated nuclear models were used to calculate reduced transition probabilities of GT transitions. Few noticeable mentions include shell model Monte Carlo technique (e.g., \cite{Dea98}), thermal quasi-particle random-phase approximation, QRPA (e.g., \cite{Mol90,Nab99, Nab04}), large-scale shell model (e.g., \cite{Lan03}), density functional theory  (e.g.,  \cite{Nom20}), Hartree-Fock-Bogoliubov method  (e.g., \cite{Eng99}) and shell model (e.g., \cite{Kum20}).

The current study  investigates the effect of pairing gaps on calculated GT strength functions and the associated $\beta$ decay rates under terrestrial and stellar conditions. The $\beta$-decay properties were studied using the quasiparticle random phase approximation model with a separable multi-shell schematic and separable interaction on top of axially symmetric-deformed mean-field calculation. Previously a similar investigation was performed separately for $sd$- \cite{Nabi22} and $fp$-shell nuclei \cite{Ull22, Ull23}. Recently a list of the top 50 electron capturing  and $\beta$ decaying nuclei, possessing the largest effect on $Y_e$ for conditions after silicon core burning till the conditions of core collapse prior to neutrino trapping,  was published \cite{Nab21}. This investigation led to the determination of most important weak interaction nuclei in the presupernova evolution of massive stars. To achieve this goal, an ensemble containing 728 nuclei in the mass range of $A$ = (1–100) was considered. The idea was to sort nuclei having the largest effect on $Y_e$ post silicon core
burning, by averaging the contribution from each nucleus to $\dot{Y_e}$ (time-rate of change of lepton fraction)
over the entire chosen stellar trajectory.
In the current project, we specifically focus on the top-ranked 50 nuclei  as per the findings of Ref.~ \cite{Nab21} with $\beta^-$ as the dominant decay mode~\cite{Aud21}, and study the effect of pairing gaps on $\beta$-decay properties of these nuclei. 

Pairing gaps are one of the most important parameters in the pn-QRPA model. It is to be noted that the present investigation includes neutron-neutron and proton-proton pairing correlations, which have only isovector contribution. For the isoscalar part, one has to include the neutron-proton ($np$) pairing correlations, not considered in the present manuscript. The current pn-QRPA model is limited as it ignores the neutron-proton $np$ pairing effect and incorporation of $np$ pairing may be taken as a future assignment. Such kind of calculations were performed earlier by author in Ref. \cite{Ha22}, albeit only for N=Z+2 nuclei. The conclusions of their study stated that isoscalar interaction behaves in a fashion similar to the tensor force interaction. In the calculations reported in Ref. \cite{Ha22} showed that the tensor force shifts the GT peak to low excitation energies. Incorporation of tension force may result in lower centroid values of calculated GT strength distributions and could lead to higher values of calculated $\beta$ decay rates. In order to compensate, the same effect of shifting calculated $\beta$ strength to lower excitation energies in the current pn-QRPA model was achieved by incorporation of particle-particle forces   (see Section 2 of Ref.~\cite{Hir93}). The pairing energy of identical nucleons in even-even isotopes can be estimated using a variety of methods based on the masses of neighboring nuclei, but despite the extensive study of the issue, the issue of which relation most closely approximates the pairing interaction is still open for debate \cite{Jen84,Sat98,Ben00,Mag23}. We chose to employ three different recipes for calculation of pairing gaps in our investigation. Details follow in the next section.

The paper is organized as follows: The theoretical framework used for calculations, is described in Section 2.  Section 3 presents the discussion on our investigation. Finally, the summary and concluding remarks of the present work are presented in Section 4.    
	\section{Formalism}
	\label{sec:2}
The Hamiltonian of the current pn-QRPA model is given as
\begin{equation} \label{eq1}
	H^{pn-QRPA} =H^{sp} + V^{pairing} + V_{GT}^{ph} + V_{GT}^{pp} ,
\end{equation}
where $H^{sp}$, $V^{pairing}$, $V_{GT}^{ph}$ and $V_{GT}^{pp}$ denote the single-particle Hamiltonian, pairing forces for BCS calculation, particle-hole ($ph$) and particle-particle ($pp$) interactions for GT strength, respectively.  The single-particle eigenfunctions and eigenvalues were computed using the Nilsson model \cite{Nil55}. Other parameters essential for the solution of Equation \ref{eq1} are nuclear deformation,  the Nilsson potential parameters (NPP),  Q-values, pairing gaps and the GT force parameters.The Q-value for $\beta^-$ decay was calculated using 
\begin{equation} \label{qval}
	Q = [m (^A_Z P) - m(^A_{Z+1}{D})  ] c^2,
\end{equation}
where $P$ is the parent nucleus, $D$ is the daughter nucleus and $m$ is the nuclear mass.

Nuclear deformation parameter ($\beta_{2}$) was determined  using the formula
\begin{equation} \label{eq2}
	\beta_{2} = \frac {125 (Q_{2})} {1.44 (A)^{2/3} (Z)},
\end{equation}
where $Q_{2}$ is the electric quadrupole moment taken from \cite{Mol16}. The NPP were chosen from \cite{Rag84}.  Nilsson oscillator constant was taken as $\hbar\omega = 41/A^{+1/3}$ in units of MeV, similar for neutrons and protons. Q-values were determined using the recent mass compilation \cite{Aud21}. 

The pairing gaps between nucleons were chosen using three different formulae. The first formula is most often used in literature~\cite{Sta90,Hir93,Hom96}. It has the same value for neutron-neutron and proton-proton pairing. It is given by
\begin{equation} \label{eq3}
	\Delta _{nn} =\Delta _{pp} =12/\sqrt{A}.
\end{equation}
This is the traditionally used formula for calculation of pairing gaps. The second formula contains three terms and is based on separation energies of neutrons and protons. It is given by 

\begin{equation} \label{eq4}
	\Delta_{nn} = \frac{1}{8} (-1)^{A-Z+1} [2S_n (A+1, Z) - 4S_n (A,Z) +2S_n(A-1, Z)]
\end{equation}
\begin{equation}
	\label{eq5}		
	\Delta_{pp} = \frac{1}{8} (-1)^{1+Z} [2S_p (A+1, Z+1) - 4S_p (A,Z) +2S_p(A-1, Z-1)].     	
\end{equation}
The third recipe contains 5 terms and is a function of the binding energies of the nucleons. It is given by 
\begin{equation} \label{eq6}
	\Delta_{nn} = \frac{1}{16}[2B(Z, N-2) - 8B(Z, N-1) + 12B(Z,N) - 8B(Z,N+1) + 2B(Z,N+2)]
\end{equation}
\begin{equation} \label{eq7}
	\Delta_{pp} = \frac{1}{16} [2B(Z-2, N) - 8B(Z-1, N) + 12B(Z,N) - 8B(Z+1,N) + 2B(Z+2,N)].
\end{equation}	
The values of binding energies were taken from Ref.~\cite{Wan21}. Henceforth in this text, we will refer to the first formula of pairing gaps as TF (one-term or traditional formula), the second as 3TF (three-term formula) and the last formula as 5TF (five-term formula). 

The spherical nucleon basis represented by ($c^{\dagger}_{jm}$, $c_{jm}$),  with angular momentum $j$ and $m$ as its z-component, was transformed into deformed basis ($d^{\dagger}_{m\alpha}$, $d_{m\alpha}$) using the transformation
\begin{equation}\label{eq8}
	~ ~ ~ ~ ~ ~ ~ ~ ~ ~ ~ ~ ~ ~ ~ ~ ~	d^{\dagger}_{m\alpha}=\Sigma_{j}D^{m\alpha}_{j}c^{\dagger}_{jm}~~~~,
\end{equation}
where $c^{\dagger}$ and $d^{\dagger}$ are the particle creation operators in the spherical and deformed basis, respectively. The transformation matrix $D$ is a set of Nilsson eigenfunctions and $\alpha$ represents the additional quantum numbers.  Later, we used the Bogoliubov transformation to introduced the quasiparticle basis $(a^{\dagger}_{m\alpha}, a_{m\alpha})$ 
\begin{equation}\label{eq9}
	~ ~ ~ ~ ~ ~ ~ ~ ~ ~ ~ ~	a^{\dagger}_{m\alpha}=u_{m\alpha}d^{\dagger}_{m\alpha}-v_{m\alpha}d_{\bar{m}\alpha}
\end{equation}
\begin{equation}\label{eq10}
	~ ~ ~ ~ ~ ~ ~ ~ ~ ~ ~ ~	a^{\dagger}_{\bar{m}\alpha}=u_{m\alpha}d^{\dagger}_{\bar{m}\alpha}+v_{m\alpha}d_{m\alpha}~~     (m > 0)  ~~~~~,    
\end{equation}
where  $\bar{m}$ represents the time-reversed state of $m$. The occupation amplitudes satisfied the condition $v^{2}_{m\alpha}$ + $u^{2}_{m\alpha}$ = 1, and were computed using the BCS equations with pairing gaps given in Equations (\ref{eq3} - \ref{eq7}). The pn-QRPA theory deals with quasiparticle states of proton-neutron systems and correlations between them. 
The ground state is a vacuum for QRPA phonon, $\hat{\Gamma}_{\omega}|$QRPA$> = 0$, with phonon creation operator defined by
\begin{eqnarray}
	\hat{\Gamma}^{\dagger}_{\omega}(\mu)&=& \sum_{\pi,\nu}X^{\pi\nu}_{\omega}(\mu)\hat{a}^{\dagger}_{\pi}\hat{a}^{\dagger}_{\bar{\nu}}-Y^{\pi\nu}_{\omega}(\mu)\hat{a}_{\nu}\hat{a}_{\bar{\pi}}~~~~~,
	\label{Eq:PCO}
\end{eqnarray}
where $\nu$ and $\pi$, respectively, denote the neutron and proton single quasiparticle states. The sum runs over all possible
$\pi\nu$-pairs which satisfy $\mu=m_{\pi}-m_{\nu}$ = (0, $\pm$1). 
The forward-going ($X_{\omega}$) and backward-going ($Y_{\omega}$) amplitudes are the eigenvectors,  whereas,  energy ($\omega$) are the eigenvalues of the
well-known (Q)RPA equation
\begin{center}
	\begin{equation}
		\left[ {\begin{array}{cc}
				M & N \\
				-N & -M \\
		\end{array} } \right]
		\left[ {\begin{array}{c}
				X \\
				Y\\
		\end{array} } \right] =  \omega \left[ {\begin{array}{c}
				X\\
				Y\\
		\end{array} } \right].
		\label{Eq:RPAM}
	\end{equation}
\end{center}
The solution of Equation~(\ref{Eq:RPAM}) was obtained for each projection value ($\mu = 0,\pm1$). Matrix elements $M$ and $N$ were determined using
\begin{eqnarray}
	M_{\pi\nu,\pi^{\prime}\nu^{\prime}}=&\delta_{\pi\nu,\pi^{\prime}\nu^{\prime}}(\varepsilon_{\pi}+\varepsilon_{\nu})\nonumber \\
	&+V^{pp}_{\pi\nu,\pi^{\prime}\nu^{\prime}}(v_{\pi}v_{\nu}v_{\pi^{\prime}}v_{\nu^{\prime}}+u_{\pi}u_{\nu}u_{\pi^{\prime}}u_{\nu^{\prime}})\nonumber \\
	&+V^{ph}_{\pi\nu,\pi^{\prime}\nu^{\prime}}(v_{\pi}u_{\nu}v_{\pi^{\prime}}u_{\nu^{\prime}}+u_{\pi}v_{\nu}u_{\pi^{\prime}}v_{\nu^{\prime}})~~~,
	\label{Mentry}
\end{eqnarray}
\begin{eqnarray}
	N_{\pi\nu,\pi^{\prime}\nu^{\prime}}=&V^{pp}_{\pi\nu,\pi^{\prime}\nu^{\prime}}(u_{\pi}u_{\nu}v_{\pi^{\prime}}v_{\nu^{\prime}}+v_{\pi}v_{\nu}u_{\pi^{\prime}}u_{\nu^{\prime}})\nonumber \\
	&-V^{ph}_{\pi\nu,\pi^{\prime}\nu^{\prime}}(v_{\pi}u_{\nu}u_{\pi^{\prime}}v_{\nu^{\prime}}+u_{\pi}v_{\nu}v_{\pi^{\prime}}u_{\nu^{\prime}})~~~,
	\label{Nentry}
\end{eqnarray}
with
\begin{equation}
	V^{pp}_{\pi\nu,\pi^{\prime}\nu^{\prime}} = V_{\pi \bar{\nu},\pi^{\prime}\bar{\nu^{\prime}}}~~~,
\end{equation}
\begin{equation}
	V^{ph}_{\pi\nu,\pi^{\prime}\nu^{\prime}} = -V_{\pi \nu^{\prime},\pi^{\prime}\nu}~~~.
\end{equation}
The quasiparticle energies ($\varepsilon_{\pi}, \varepsilon_{\nu}$) were obtained from the BCS calculations.
We used separable GT residual forces, namely:
particle-hole ($ph$) and particle-particle ($pp$) forces in our calculation. We took $pp$ GT force as
\begin{equation}
	V_{GT}^{pp} = -2\kappa\sum_{\mu} (-1)^{\mu}\hat{P}^{\dagger}_{\mu}\hat{P}_{-\mu}~~~,
	\label{ppGT}
\end{equation}
where
\begin{equation}
	\hat{P}^{\dagger}_{\mu} = \sum_{j_{\pi}m_{\pi}{j_{\nu}m_{\nu}}} \langle j_{\nu}m_{\nu}|(\sigma_{\mu}\tau_{-})^{\dagger}|j_{\pi}m_{\pi} \rangle (-1)^{l_{\nu}+j_{\nu}-m_{\nu}}\hat{c}^{\dagger}_{j_{\pi}m_{\pi}}\hat{c}^{\dagger}_{j_{\nu}-m_{\nu}}~~~,
\end{equation}
and $ph$ GT force as
\begin{equation}
	V_{GT}^{ph} = 2\chi\sum_{\mu} (-1)^{\mu}\hat{R}_{\mu}\hat{R}^{\dagger}_{-\mu}~~~,
	\label{phGT}
\end{equation}
where
\begin{equation}
	\hat{R}_{\mu} = \sum_{j_{\pi}m_{\pi}{j_{\nu}m_{\nu}}} \langle j_{\pi}m_{\pi}|\sigma_{\mu}\tau_{-}|j_{\nu}m_{\nu} \rangle\hat{c}^{\dagger}_{j_{\pi}m_{\pi}}\hat{c}_{j_{\nu}m_{\nu}}~~~,
\end{equation}
where $\chi$ and $\kappa$ are the $pp$ and $ph$ GT force parameters, respectively. With the use of the separable GT forces in our calculation, the RPA matrix equation reduced to a 4$^{th}$ order algebraic equation. The method to determine the roots of these equations can be seen from~\cite{Mut92}. This simplification saved the computational time when compared to the full diagonalization of the
nuclear Hamiltonian (Equation~\ref{eq1}).

In the RPA formalism, excitations from the ground state ($J^{\pi} = 0^+$) of an even-even nucleus is considered.
The ground-state of an odd-odd (odd-A) parent nucleus is expressed as a proton-neutron quasiparticle pair
(one quasiparticle) state of the smallest energy. Then two possible transitions are the phonon excitations (in which
quasiparticle merely plays the role of a spectator) and transition of the quasiparticle itself. In
the latter case, correlations of phonon to the quasiparticle transitions were treated in first-order perturbation~\cite{Mol90}.

We next present quasiparticle transitions, construction of phonon-related multi-quasiparticle states (representing nuclear excited levels of even-even, odd-A and odd-odd nuclei) and formulae
of GT transitions within the current model using the recipe given by~\cite{Mut92}. It is again reminded that the occupation amplitudes of the quasiparticle states were calculated within BCS formalism using three different pairing gap values. 
The phonon-correlated one quasiparticle states were defined by
\begin{eqnarray}
	|\pi_{corr}\rangle~=~a^\dagger_{\pi}|-\rangle +& \sum_{\nu, \omega}a^\dagger_{\nu}A^\dagger_{\omega}(\mu)|-\rangle \nonumber~\langle-|[a^\dagger_{\nu}A^\dagger_{\omega}(\mu)]^{\dagger}H_{31}a^\dagger_{\pi}|-\rangle \nonumber \\
	&\times E_{\pi}(\nu,\omega)~~~,
	\label{opn1}
\end{eqnarray}
\begin{eqnarray}
	|\nu_{corr}\rangle~=~a^\dagger_{\nu}|-\rangle +& \sum_{\pi, \omega}a^\dagger_{\pi}A^\dagger_{\omega}(-\mu)|-\rangle \nonumber~\langle-|[a^\dagger_{\pi}A^\dagger_{\omega}(-\mu)]^{\dagger}H_{31}a^\dagger_{\nu}|-\rangle \nonumber \\
	&\times E_{\nu}(\pi,\omega)~~~,
	\label{opn2}
\end{eqnarray}
with
\begin{equation}
	E_{a}(b,\omega)=\frac{1}{\epsilon_{a}-\epsilon_{b}-\omega}~~~~~~~a, b = \pi, \nu ~~~~,
	\label{opn3}
\end{equation}
and
\begin{eqnarray}
	H_{31}=\sum V_{\pi \nu,\bar{\pi^{\prime}}\nu^{\prime}}(u_{\pi}u_{\nu}v_{\pi^{\prime}}u_{\nu^{\prime}}-v_{\pi}v_{\nu}u_{\pi^{\prime}}v_{\nu^{\prime}}) (a^\dagger_{\pi}a^\dagger_{\nu}a^\dagger_{\pi^{\prime}}a_{\nu^{\prime}} + h.c.) + \nonumber \\
	\sum V_{\pi \nu,\pi^{\prime}\bar{\nu^{\prime}}}(v_{\pi}v_{\nu}v_{\pi^{\prime}}u_{\nu^{\prime}}-u_{\pi}u_{\nu}u_{\pi^{\prime}}v_{\nu^{\prime}}) (a^\dagger_{\pi}a^\dagger_{\nu}a^\dagger_{\nu^{\prime}}a_{\pi^{\prime}} + h.c.) ~~~,
	\label{opn3}
\end{eqnarray} 
where $h.c.$ stands for Hermitian conjugate. The terms $E_{a}(b,\omega)$ can be modified to prevent the singularity in the transition amplitude
caused by the first-order perturbation of the odd-particle wave function. The first term in Equation~(\ref{opn1}) and Equation~(\ref{opn2}) denotes the proton (neutron) quasiparticle state, while the second term denotes RPA correlated
phonons admixed with quasiparticle phonon coupled Hamiltonian H$_{31}$, which was accomplished by Bogoliubov
transformation from separable $pp$ and $ph$ GT interaction forces. The summation applies to all phonon states and
neutron (proton) quasiparticle states, satisfying $m_{\pi}-m_{\nu}=\mu$ with $\pi_{\pi}\pi_{\nu}=1$. Calculation of the quasiparticle transition amplitudes for correlated states
can be seen from~\cite{Mut89}. The amplitudes of GT transitions in terms of separable forces are
\begin{eqnarray}
	<{\pi_{corr}}|\tau_-\sigma_{\mu}|{\nu_{corr}}>~=~ q^U_{\pi\nu}+ 2\chi [q^U_{\pi\nu}\sum\limits_{\omega}(Z^{-2}_\omega E_\pi(\nu,\omega)+Z^{+2}_{\omega}E_\nu(\pi,\omega)) \nonumber\\
	-q^V_{\pi\nu}\sum\limits_{\omega}Z^-_{\omega}Z^+_{\omega}(E_\pi(\nu,\omega)+E_\nu(\pi,\omega))]
	+2\kappa[q_{\pi\nu}\sum\limits_{\omega}(Z^-_{\omega}Z^{--}_{\omega}E_\pi(\nu,\omega)-Z^+_{\omega}Z^{++}_{\omega}E_\nu(\pi,\omega)) \nonumber\\
	-\tilde{q}_{\pi\nu}\sum\limits_{\omega}(Z^-_{\omega}Z^{++}_{\omega}E_\pi(\nu,\omega)-Z^+_{\omega}Z^{--}_{\omega}E_\nu(\pi,\omega))],
	\label{opn4}	
\end{eqnarray}
\begin{eqnarray}
	%	\begin{split}
		<{\pi_{corr}}|\tau_+\sigma_{\mu}|{\nu_{corr}}>=q^V_{\pi\nu}+2\chi[q^V_{\pi\nu}\sum\limits_{\omega}(Z^{+2}_{\omega}E_\pi(\nu,\omega)+Z^{-2}_{\omega}E_\nu(\pi,\omega)) \nonumber\\
		-q^U_{\pi\nu}\sum\limits_{\omega}Z^-_{\omega}Z^+_{\omega}(E_\pi(\nu,\omega)+E_\nu(\pi,\omega))]
		+2\kappa[\tilde{q}_{\pi\nu}\sum\limits_{\omega}(Z^+_{\omega}Z^{++}_{\omega}E_\pi(\nu,\omega)
		-Z^-_{\omega}Z^{--}_{\omega}E_\nu(\pi,\omega))\nonumber\\-q_{\pi\nu}\sum\limits_{\omega}(Z^+_{\omega}Z^{--}_{\omega}E_\pi(\nu,\omega)-Z^-_{\omega}Z^{++}_{\omega}E_\nu(\pi,\omega))],
		%	\end{split}
	\label{opn5}
\end{eqnarray}
\begin{equation}
	<{\nu_{corr}}|\tau_{\pm}\sigma_{-\mu}|{\pi_{corr}}>=(-1)^{\mu}<{\pi_{corr}}|\tau_{\mp}\sigma_{\mu}|{\nu_{corr}}>.
	\label{opn6}
\end{equation}
In Eqs.~(\ref{opn4}), (\ref{opn5}) and (\ref{opn6}), $\sigma_{\mu}$ and $\tau_{\pm}$ are spin and iso-spin type operators,  respectively, and other symbols $q_{\pi\nu}$ ($\tilde{q}_{\pi\nu}$), $q^U_{\pi\nu}$ ($q^V_{\pi\nu}$), $Z^-_{\omega}$ ($Z^+_{\omega}$) and $Z^{--}_{\omega}$ ($Z^{++}_{\omega}$) are defined as
\begin{eqnarray}
	q_{\pi\nu}=f_{\pi\nu}u_\pi v_\nu,~~~~ q_{\pi\nu}^{U}=f_{\pi\nu}u_\pi u_\nu, \nonumber \\
	\tilde q_{\pi\nu}=f_{\pi\nu}v_\pi u_\nu,~~~~_{\pi\nu}^{V}=f_{\pi\nu}v_\pi v_\nu \nonumber \\
	Z^{-}_{\omega}= \sum_{\pi,\nu}(X^{\pi\nu}_{\omega}q_{\pi\nu}-Y^{\pi\nu}_{\omega}\tilde q_{\pi\nu}),\nonumber \\
	Z^{+}_{\omega}= \sum_{\pi,\nu}(X^{\pi\nu}_{\omega}\tilde q_{\pi\nu}-Y^{\pi\nu}_{\omega}q_{\pi\nu}),
	\nonumber \\
	Z^{--}_{\omega}= \sum_{\pi,\nu}(X^{\pi\nu}_{\omega}q^{U}_{\pi\nu}+Y^{\pi\nu}_{\omega}q^{V}_{\pi\nu}),
	\nonumber \\
	Z^{+ +}_{\omega}=
	\sum_{\pi\nu}(X^{\pi,\nu}_{\omega}q^{V}_{\pi\nu}+Y^{\pi\nu}_{\omega}q^{U}_{\pi\nu}).
\end{eqnarray}
The terms $X^{\pi\nu}_{\omega}$ and $Y^{\pi\nu}_{\omega}$ were defined earlier and other symbols have usual meanings.
The idea of quasiparticle transitions with first-order phonon correlations can be extended to an odd-odd parent nucleus.
The ground state is assumed to be a proton-neutron quasiparticle pair state of the smallest energy. The GT transitions
of the quasiparticle lead to two-proton or two-neutron quasiparticle states in the even-even daughter nucleus. The
two quasiparticle states were constructed with phonon correlations and given by
\begin{eqnarray}
	|{\pi \nu_{corr}}>~=~a_\pi^\dagger a^\dagger_\nu|{-}>+\frac{1}{2}\sum\limits_{\pi'_1,\pi'_2,\omega}a^\dagger_{\pi'_1}a^\dagger_{\pi'_2}A^\dagger_{\omega}(-\mu)|{-}>~~~~~~~~~~~~~~~\nonumber\\\times <{-}|[a^\dagger_{\pi'_1}a^\dagger_{\pi'_2}A^\dagger_{\omega}(-\mu)]^\dagger H_{31}a^\dagger_\pi a^\dagger_\nu|{-}>E_{\pi\nu}(\pi'_1\pi'_2,\omega)+\frac{1}{2}\sum\limits_{\nu'_1,\nu'_2,\omega}a^\dagger_{\nu'_1}a_{\nu'_2}A^\dagger_{\omega}(\mu)|{-}>\nonumber\\\times <{-}|[a^\dagger_{\nu'_1}a^\dagger_{\nu'_2}A^\dagger_{\omega}(\mu)]^\dagger
	H_{31}a^\dagger_\pi a^\dagger_\nu|{-}>E_{\pi\nu}(\nu'_1\nu'_2,\omega)~~~,
	\label{opn7}
\end{eqnarray}
\begin{eqnarray}
	%	\begin{split}
		<{\pi_1\pi_{2corr}}|~=~a^\dagger_{\pi_1}a^\dagger_{\pi_2}|{-}>+\sum\limits_{\pi',\nu',\omega}a^\dagger_{\pi'}a^\dagger_{\nu'}A^\dagger_{\omega}(\mu)|{-}>~~~~~~~~~~~~~~~\nonumber\\\times <{-}|[a^\dagger_{\pi'}a^\dagger_{\nu'}A^\dagger_{\omega}(\mu)]^\dagger
		H_{31}a^\dagger_{\pi_1}a^\dagger_{\pi_2}|{-}>E_{\pi_1\pi_2}(\pi'\nu',\omega)~~~,
		\label{opn8}
		%	\end{split}
\end{eqnarray}
\begin{eqnarray}
	%	\begin{split}
		<{\nu_1\nu_{2corr}}|~=~a^\dagger_{\nu_1}a^\dagger_{\nu_2}|{-}>+\sum\limits_{\pi',\nu',\omega}a^+_{\pi'}a^\dagger_{\nu'}A^\dagger_{\omega}(-\mu)|{-}>~~~~~~~~~~~~~~~\nonumber\\\times <{-}|[a^\dagger_{\pi'}a^\dagger_{\nu'}A^\dagger_{\omega}(-\mu)]^\dagger
		H_{31}a^\dagger_{\nu_1}a^\dagger_{\nu_2}|{-}>E_{\nu_1\nu_2}(\pi'\nu',\omega)~~~,
		\label{opn9}
	\end{eqnarray}
	where
	\begin{equation}
		E_{ab}(cd,\omega)=\frac{1}{(\epsilon_a+\epsilon_b)-(\epsilon_{c}+\epsilon_{d}+\omega)}~~~<
		\label{opn10}
	\end{equation}
	where subscript index a (b) denotes $\pi,~\pi_1$ and $\nu_1$ ($\nu,~\pi_2$ and $\nu_2$) and c (d) denotes
	$\pi',~\pi'_1$ and $\nu'_1$ ($\nu',~\pi'_2$ and $\nu'_2$).
	The GT transition amplitudes between these states were reduced to those of one quasiparticle states
	\begin{eqnarray}
		<{\pi_1\pi_{2corr}}|\tau_{\pm}\sigma_{\mu}|{\pi \nu_{corr}}>~=~&\delta(\pi_1,\pi)<{\pi_{2corr}}|\tau_{\pm}\sigma_{\mu}|{\nu_{corr}}>\nonumber\\ &-\delta(\pi_2,\pi)
		<{\pi_{1corr}}|\tau_{\pm}\sigma_{\mu}|{\nu_{corr}}>~~~,
		\label{opn11}
	\end{eqnarray}
	\begin{eqnarray}
		<{\nu_1\nu_{2corr}}|\tau_{\pm}\sigma_{-\mu}|{\pi \nu_{corr}}>~=~&\delta(\nu_2,\nu)<{\nu_{1corr}}|\tau_{\pm}\sigma_{-\mu}|{\pi_{corr}}>\nonumber\\ &-\delta(\nu_1,\nu)
		<{\nu_{2corr}}|\tau_{\pm}\sigma_{-\mu}|{\pi_{corr}}>~~~,
		\label{opn12}
	\end{eqnarray}
	by ignoring terms of second order in the correlated phonons.
	For odd-odd parent nuclei, QRPA phonon excitations are also possible where the quasiparticle pair acts as spectators in the same single quasiparticle shells. The nuclear excited states can be constructed as phonon correlated multi quasiparticle states. The transition amplitudes between the multi quasiparticle states can be reduced to those of one
	quasiparticle states as described below.
	
	Excited levels of an even-even nucleus are two-proton quasiparticle states and two-neutron quasiparticle states.
	Transitions from these initial states to final neutron-proton quasiparticle pair states are possible
	in the odd-odd daughter nuclei. The transition amplitudes can be reduced to correlated quasiparticle states
	by taking the Hermitian conjugate of Eq.~(\ref{opn11}) and (\ref{opn12})
	\begin{eqnarray}
		<{\pi \nu_{corr}}|\tau_{\pm}\sigma_{-\mu}|{\pi_1\pi_{2corr}}>~=~& - \delta(\pi,\pi_2)<{\nu_{corr}}|\tau_{\pm}\sigma_{-\mu}|{\pi_{1corr}}>\nonumber\\&+\delta(\pi,\pi_1)
		<{\nu_{corr}}|\tau_{\pm}\sigma_{-\mu}|{\pi_{2corr}}>~~~,
		\label{opn13}
	\end{eqnarray}
	\begin{eqnarray}
		<{\pi \nu_{corr}}|\tau_{\pm}\sigma_{\mu}|{\nu_1\nu_{2corr}}>~=~&\delta(\nu,\nu_2)<{\pi_{corr}}|\tau_{\pm}\sigma_{\mu}|{\nu_{1corr}}>\nonumber\\&-\delta(\nu,\nu_1)
		<{\pi_{corr}}|\tau_{\pm}\sigma_{\mu}|{\nu_{2corr}}>~~~.
		\label{opn13}
	\end{eqnarray}
	
	When a nucleus has an odd nucleon (a proton and/or a neutron), low-lying states were obtained by lifting
	the quasiparticle in the orbit of the smallest energy to higher-lying orbits. States of an odd-proton even-neutron nucleus
	were expressed by three-proton states or one proton two-neutron states, corresponding to excitation
	of a proton or a neutron
	\begin{eqnarray}\label{opn14}
		|\pi_1\pi_2\pi_{3corr}\rangle~=~a^\dagger_{\pi_1}a^\dagger_{\pi_2}a^\dagger_{\pi_3}|-\rangle + \frac{1}{2}\sum_{\pi^{'}_1,\pi^{'}_2,\nu^{'},\omega}a^\dagger_{\pi^{'}_1}a^\dagger_{\pi^{'}_2}a^\dagger_{\nu^{'}}A^\dagger_{\omega}(\mu)|-\rangle \nonumber \\
		~~~~~~~~~~~~~~~ \times \langle-|[a^\dagger_{\pi^{'}_1}a^\dagger_{\pi^{'}_2}a^\dagger_{\nu^{'}}A^\dagger_{\omega}(\mu)]^{\dagger}H_{31}a^\dagger_{\pi_1}a^\dagger_{\pi_2}a^\dagger_{\pi_3}|-\rangle E_{\pi_1\pi_2\pi_3}(\pi^{'}_1\pi^{'}_2\nu^{'},\omega)~~~,
	\end{eqnarray}
	\begin{eqnarray}\label{opn15}
		|\pi_1\nu_1\nu_{2corr}\rangle ~=~a^\dagger_{\pi_1}a^\dagger_{\nu_1}a^\dagger_{\nu_2}|-\rangle \nonumber + \frac{1}{2}\sum_{\pi^{'}_1,\pi^{'}_2,\nu^{'},\omega}a^\dagger_{\pi^{'}_1}a^\dagger_{\pi^{'}_2}a^\dagger_{\nu^{'}}A^\dagger_{\omega}(-\mu)|-\rangle \nonumber \\
		~~~~~~~~~~~~~~~ \times \langle-|[a^\dagger_{\pi^{'}_1}a^\dagger_{\pi^{'}_2}a^\dagger_{\nu^{'}}A^\dagger_{\omega}(-\mu)]^{\dagger}H_{31}a^\dagger_{\pi_1}a^\dagger_{\nu_1}a^\dagger_{\nu_2}|-\rangle \nonumber \\
		~~~~~~~~~~~~~~~ \times E_{\pi_1\nu_1\nu_2}(\pi^{'}_1\pi^{'}_2\nu^{'},\omega) \nonumber +\frac{1}{6}\sum_{\nu^{'}_1,\nu^{'}_2,\nu^{'}_3,\omega}a^\dagger_{\nu^{'}_1}a^\dagger_{\nu^{'}_2}a^\dagger_{\nu^{'}_3}A^\dagger_{\omega}(\mu)|-\rangle \nonumber \\
		~~~~~~~~~~~~~~~ \times \langle-|[a^\dagger_{\nu^{'}_1}a^\dagger_{\nu^{'}_2}a^\dagger_{\nu^{'}_3}A^\dagger_{\omega}(\mu)]^{\dagger}H_{31}a^\dagger_{\pi_1}a^\dagger_{\nu_1}a^\dagger_{\nu_2}|-\rangle E_{\pi_1\nu_1\nu_2}(\nu^{'}_1\nu^{'}_2\nu^{'}_3,\omega)~~~,
	\end{eqnarray}
	with the energy denominators of first order perturbation
	\begin{equation}
		E_{abc}(def,\omega)=\frac{1}{(\epsilon_{a}+\epsilon_{b}+\epsilon_{c}-\epsilon_{d}-\epsilon_{e}-\epsilon_{f}-\omega)}~~~,
		\label{opn16}
	\end{equation}
	where subscripts represent $\pi_1$, $\pi_2$, $\pi_3$, $\pi$, $\nu_1$ and $\nu_2$ ($\pi'_1$, $\pi'_2$, $\nu'$, $\nu'_1$, $\nu'_2$ and $\nu'_2$). These equations can be used to generate the three quasiparticle states of odd-proton and even-neutron by swapping the neutron and proton states, $\nu\longleftrightarrow \pi$ and $A^{\dagger}_\omega(\mu) \longleftrightarrow A^{\dagger}_\omega(-\mu)$. Amplitudes of the quasiparticle transitions between the three quasiparticle states were reduced to those for correlated one quasiparticle states. For parent nuclei with an odd proton
	\begin{eqnarray}
		\langle \pi^{'}_1\pi^{'}_2\nu^{'}_{1corr}|\tau_{\pm}\sigma_{-\mu}|\pi_1\pi_2\pi_{3corr}\rangle&
		~=~\delta(\pi^{'}_1,\pi_2)\delta(\pi^{'}_2,\pi_3)\langle \nu^{'}_{1corr}|\tau_{\pm}\sigma_{-\mu}|\pi_{1corr}\rangle \nonumber\\
		~&-\delta(\pi^{'}_1,\pi_1)\delta(\pi^{'}_2,\pi_3)\langle \nu^{'}_{1corr}|\tau_{\pm}\sigma_{-\mu}|\pi_{2corr}\rangle \nonumber\\
		~&+\delta(\pi^{'}_1,\pi_1)\delta(\pi^{'}_2,\pi_2)\langle \nu^{'}_{1corr}|\tau_{\pm}\sigma_{-\mu}|\pi_{3corr}\rangle~~~,
	\end{eqnarray}
	\begin{eqnarray}
		\langle \pi^{'}_1\pi^{'}_2\nu^{'}_{1corr}|\tau_{\pm}\sigma_{\mu}|\pi_1\nu_1\nu_{2corr}\rangle&
		~=~\delta(\nu^{'}_1,\nu_2)[\delta(\pi^{'}_1,\pi_1)\langle \pi^{'}_{2corr}|\tau_{\pm}\sigma_{\mu}|\nu_{1corr}\rangle \nonumber\\
		~&-\delta(\pi^{'}_2,\pi_1)\langle \pi^{'}_{1corr}|\tau_{\pm}\sigma_{\mu}|\nu_{1corr}\rangle] \nonumber\\
		~&-\delta(\nu^{'}_1,\nu_1)[\delta(\pi^{'}_1,\pi_1)\langle \pi^{'}_{2corr}|\tau_{\pm}\sigma_{\mu}|\nu_{2corr}\rangle \nonumber\\
		~&-\delta(\pi^{'}_2,\pi_1)\langle \pi^{'}_{1corr}|\tau_{\pm}\sigma_{\mu}|\nu_{2corr}\rangle]~~~,
	\end{eqnarray}
	\begin{eqnarray}
		\langle \nu^{'}_1\nu^{'}_2\nu^{'}_{3corr}|\tau_{\pm}\sigma_{-\mu}|\pi_1\nu_1\nu_{2corr}\rangle &
		~=~\delta(\nu^{'}_2,\nu_1)\delta(\nu^{'}_3,\nu_2)\langle \nu^{'}_{1corr}|\tau_{\pm}\sigma_{-\mu}|\pi_{1corr}\rangle \nonumber\\
		~&-\delta(\nu^{'}_1,\nu_1)\delta(\nu^{'}_3,\nu_2)\langle \nu^{'}_{2corr}|\tau_{\pm}\sigma_{-\mu}|\pi_{1corr}\rangle \nonumber\\
		~&+\delta(\nu^{'}_1,\nu_1)\delta(\nu^{'}_2,\nu_2)\langle \nu^{'}_{3corr}|\tau_{\pm}\sigma_{-\mu}|\pi_{1corr}\rangle~~~,
	\end{eqnarray}
	and for parent nuclei with an odd neutron
	\begin{eqnarray}
		\langle \pi^{'}_1\nu^{'}_1\nu^{'}_{2corr}|\tau_{\pm}\sigma_{\mu}|\nu_1\nu_2\nu_{3corr}\rangle
		&~=~\delta(\nu^{'}_1,\nu_2)\delta(\nu^{'}_2,\nu_3)\langle \pi^{'}_{1corr}|\tau_{\pm}\sigma_{\mu}|\nu_{1corr}\rangle \nonumber\\
		~&-\delta(\nu^{'}_1,\nu_1)\delta(\nu^{'}_2,\nu_3)\langle \pi^{'}_{1corr}|\tau_{\pm}\sigma_{\mu}|\nu_{2corr}\rangle \nonumber\\
		~&+\delta(\nu^{'}_1,\nu_1)\delta(\nu^{'}_2,\nu_2)\langle \pi^{'}_{1corr}|\tau_{\pm}\sigma_{\mu}|\nu_{3corr}\rangle~~~,
	\end{eqnarray}
	\begin{eqnarray}
		\langle \pi^{'}_1\nu^{'}_1\nu^{'}_{2corr}|\tau_{\pm}\sigma_{-\mu}|\pi_1\pi_2\nu_{1corr}\rangle
		&~=~\delta(\pi^{'}_1,\pi_2)[\delta(\nu^{'}_1,\nu_1)\langle \nu^{'}_{2corr}|\tau_{\pm}\sigma_{-\mu}|\pi_{1corr}\rangle \nonumber\\
		&~-\delta(\nu^{'}_2,\nu_1)\langle \nu^{'}_{1corr}|\tau_{\pm}\sigma_{-\mu}|\pi_{1corr}\rangle] \nonumber\\
		~&-\delta(\pi^{'}_1,\pi_1)[\delta(\nu^{'}_1,\nu_1)\langle \nu^{'}_{2corr}|\tau_{\pm}\sigma_{-\mu}|\pi_{2corr}\rangle \nonumber\\
		~&-\delta(\nu^{'}_2,\nu_1)\langle \nu^{'}_{1corr}|\tau_{\pm}\sigma_{-\mu}|\pi_{2corr}\rangle]~~~,
	\end{eqnarray}
	\begin{eqnarray}
		\langle \pi^{'}_1\pi^{'}_2\pi^{'}_{3corr}|\tau_{\pm}\sigma_{\mu}|\pi_1\pi_2\nu_{1corr}\rangle
		&~=~\delta(\pi^{'}_2,\pi_1)\delta(\pi^{'}_3,\pi_2)\langle \pi^{'}_{1corr}|\tau_{\pm}\sigma_{\mu}|\nu_{1corr}\rangle \nonumber\\
		&~-\delta(\pi^{'}_1,\pi_1)\delta(\pi^{'}_3,\pi_2)\langle \pi^{'}_{2corr}|\tau_{\pm}\sigma_{\mu}|\nu_{1corr}\rangle \nonumber\\
		~&+\delta(\pi^{'}_1,\pi_1)\delta(\pi^{'}_2,\pi_2)\langle \pi^{'}_{3corr}|\tau_{\pm}\sigma_{\mu}|\nu_{1corr}\rangle~~~.
	\end{eqnarray}
	
	Low-lying states in an odd-odd nucleus were expressed
	in the quasiparticle picture by proton-neutron pair states (two quasiparticle states) or by states which were
	obtained by adding two proton or two-neutron quasi-particles (four quasiparticle states). Transitions from the former
	states were described earlier. Phonon-correlated four quasiparticle states can be constructed
	similarly to the two and three quasiparticle states. Also in this
	case, transition amplitudes for the four quasiparticle states were
	reduced into those for the correlated one quasiparticle states	
	\begin{eqnarray}
		<{\pi^{'}_1\pi^{'}_2\nu^{'}_1\nu^{'}_{2corr}}|\tau_{\pm}\sigma_{-\mu}|{\pi_1\pi_2\pi_3\nu_{1corr}}
		>~~~~~~~~~~~~~~~~~~~~~~~~~~~~~~~~~~~~~~~~~~~~~~~~~~~~~~~~~~~~~~~~~~~~~~~\nonumber\\~~~~~~~~~~~~~~=~\delta(\nu^{'}_2,\nu_1)[\delta(\pi^{'}_1,\pi_2)\delta(\pi^{'}_2,\pi_3)<{\nu^{'}_{1corr}}|\tau_{\pm}\sigma_{-\mu}|{\pi_{1corr}}>\nonumber\\
		-\delta(\pi^{'}_1,\pi_1)\delta(\pi^{'}_2,\pi_3)<{\nu^{'}_{1corr}}|\tau_{\pm}\sigma_{-\mu}|{\pi_{2corr}}>\nonumber\\+\delta(\pi^{'}_1,\pi_1)\delta(\pi^{'}_2,\pi_2)<{\nu^{'}_{1corr}}|\tau_{\pm}\sigma_{-\mu}|{\pi_{3corr}}>]\nonumber\\
		-\delta(\nu^{'}_1,\nu_1)[\delta(\pi^{'}_1,\pi_2)\delta(\pi^{'}_2,\pi_3)<{\nu^{'}_{2corr}}|\tau_{\pm}\sigma_{-\mu}|{\pi_{1corr}}>\nonumber\\
		-\delta(\pi^{'}_1,\pi_1)\delta(\pi^{'}_2,\pi_3)<{\nu^{'}_{2corr}}|\tau_{\pm}\sigma_{-\mu}|{\pi_{2corr}}>\nonumber\\
		+\delta(\pi^{'}_1,\pi_1)\delta(\pi^{'}_2,\pi_2)<{\nu^{'}_{2corr}}|\tau_{\pm}\sigma_{-\mu}|{\pi_{3corr}}>]~~~,
	\end{eqnarray}
	\begin{eqnarray}
		<{\pi^{'}_1\pi^{'}_2\pi^{'}_3\pi^{'}_{4corr}}|\tau_{\pm}\sigma_{\mu}|{\pi_1\pi_2\pi_3\nu_{1corr}}>~~~~~~~~~~~~~~~~~~~~~~~~~~~~~~~~~~~~~~~~~~~~~~~~~~~~~~~~~~~~~~~~~~~~~~~\nonumber\\~~~~~~~~~~~~~~~=~~-\delta(\pi^{'}_2,\pi_1)\delta(\pi^{'}_3,\pi_2)\delta(\pi^{'}_4,\pi_3)<{\pi^{'}_{1corr}}|\tau_{\pm}\sigma_{\mu}|{\nu_{1corr}}>\nonumber\\
		+\delta(\pi^{'}_1,\pi_1)\delta(\pi^{'}_3,\pi_2)\delta(\pi^{'}_4,\pi_3)<{\pi^{'}_{2corr}}|\tau_{\pm}\sigma_{\mu}|{\nu_{1corr}}>\nonumber\\
		-\delta(\pi^{'}_1,\pi_1)\delta(\pi^{'}_2,\pi_2)\delta(\pi^{'}_4,\pi_3)<{\pi^{'}_{3corr}}|\tau_{\pm}\sigma_{\mu}|{\nu_{1corr}}>\nonumber\\
		+\delta(\pi^{'}_1,\pi_1)\delta(\pi^{'}_2,\pi_2)\delta(\pi^{'}_3,\pi_3)<{\pi^{'}_{4corr}}|\tau_{\pm}\sigma_{\mu}|{\nu_{1corr}}>~~~,
	\end{eqnarray}
	\begin{eqnarray}
		<{\pi^{'}_1\pi^{'}_2\nu^{'}_1\nu^{'}_{2corr}}|\tau_{\pm}\sigma_{\mu}|{\pi_1\nu_1\nu_2\nu_{3corr}}>~~~~~~~~~~~~~~~~~~~~~~~~~~~~~~~~~~~~~~~~~~~~~~~~~~~~~~~~~~~~~~~~~~~~~~~\nonumber\\~~~~~~~~~~~~~~~=~~\delta(\pi^{'}_1,\pi_1)[\delta(\nu^{'}_1,\nu_2)\delta(\nu^{'}_2,\nu_3)<{\pi^{'}_{2corr}}|\tau_{\pm}\sigma_{\mu}|{\nu_{1corr}}>\nonumber\\
		-\delta(\nu^{'}_1,\nu_1)\delta(\nu^{'}_2,\nu_3)<{\pi^{'}_{2corr}}|\tau_{\pm}\sigma_{\mu}|{\nu_{2corr}}>\nonumber\\
		+\delta(\nu^{'}_1,\nu_1)\delta(\nu^{'}_2,\nu_2)<{\pi^{'}_{2corr}}|\tau_{\pm}\sigma_{\mu}|{\nu_{3corr}}>]\nonumber\\
		-\delta(\pi^{'}_2,\pi_1)[\delta(\nu^{'}_1,\nu_2)\delta(\nu^{'}_2,\nu_3)<{\pi^{'}_{1corr}}|\tau_{\pm}\sigma_{\mu}|{\nu_{1corr}}>\nonumber\\
		-\delta(\nu^{'}_1,\nu_1)\delta(\nu^{'}_2,\nu_3)<{\pi^{'}_{1corr}}|\tau_{\pm}\sigma_{\mu}|{\nu_{2corr}}>\nonumber\\
		+\delta(\nu^{'}_1,\nu_1)\delta(\nu^{'}_2,\nu_2)<{\pi^{'}_{1corr}}|\tau_{\pm}\sigma_{\mu}|{\nu_{3corr}}>]~~~,
	\end{eqnarray}
	\begin{eqnarray}
		<{\nu^{'}_1\nu^{'}_2\nu^{'}_3\nu^{'}_{4corr}}|\tau_{\pm}\sigma_{-\mu}|{\pi_1\nu_1\nu_2\nu_{3corr}}>~~~~~~~~~~~~~~~~~~~~~~~~~~~~~~~~~~~~~~~~~~~~~~~~~~~~~~~~~~~~~~~~~~~~~~~\nonumber\\~~~~~~~~~~~~~~~=~~+\delta(\nu^{'}_2,\nu_1)\delta(\nu^{'}_3,\nu_2)\delta(\nu^{'}_4,\nu_3)<{\nu^{'}_{1corr}}|\tau_{\pm}\sigma_{-\mu}|{\pi_{1corr}}>\nonumber\\
		-\delta(\nu^{'}_1,\nu_1)\delta(\nu^{'}_3,\nu_2)\delta(\nu^{'}_4,\nu_3)<{\nu^{'}_{2corr}}|\tau_{\pm}\sigma_{-\mu}|{\pi_{1corr}}>\nonumber\\
		+\delta(\nu^{'}_1,\nu_1)\delta(\nu^{'}_2,\nu_2)\delta(\nu^{'}_4,\nu_3)<{\nu^{'}_{3corr}}|\tau_{\pm}\sigma_{-\mu}|{\pi_{1corr}}>\nonumber\\
		-\delta(\nu^{'}_1,\nu_1)\delta(\nu^{'}_2,\nu_2)\delta(\nu^{'}_3,\nu_3)<{\nu^{'}_{4corr}}|\tau_{\pm}\sigma_{-\mu}|{\pi_{1corr}}>~~~.
	\end{eqnarray}
	The antisymmetrization of the quasi-particles was duly taken into account for each of these amplitudes.\\
	$\pi^{'}_4>\pi^{'}_3>\pi^{'}_2>\pi^{'}_1$,~~ $\nu^{'}_4>\nu^{'}_3>\nu^{'}_2>\nu^{'}_1$,~~  $\pi_4>\pi_3>\pi_2>\pi_1$,~~ $\nu_4>\nu_3>\nu_2>\nu_1$.\\
	The GT transitions were taken into account for each phonon's excited state. It was assumed that the quasiparticle in the parent nucleus occupied the same orbit as the excited phonons.
	
	The $\beta$ decay partial half-lives $t_{1/2}$ from parent ground state were calculated using the relation
	\begin{eqnarray} \label{eq18}
		t_{p(1/2)} = \frac{D}{f_V(Z, E, A)B_F(\omega)+(g_V/g_A)^{-2}f_A(Z, E, A)B_{GT}(\omega)}~~~,
	\end{eqnarray}
	where  $E$ = ($Q$ - $\omega$).  The integrals of the available phase space for axial vector and vector transitions are denoted as $f_A(Z, A, E)$ and $f_V(Z, A, E)$, respectively. The total $\beta$-decay half-lives were computed,  including all transition probabilities to the states in the daughter within the $Q$ window.
	
	The stellar $\beta$ decay rates from the \textit{n$^{th}$} parent state to \textit{m$^{th}$} daughter level was calculated using
	\begin{equation} \label{eq9}
		\lambda _{nm}^{\beta} =\ln 2\frac{f_{nm}(T,\rho
			,E_{f})}{(ft)_{nm}}.
	\end{equation}
	The term $(ft)_{nm}$ is linked to the reduced transition probabilities ($B_{nm}$) of Fermi and GT transitions
	\begin{equation} \label{eq10}
		(ft)_{nm} =D/B_{nm},
	\end{equation}
	where
	\begin{equation} \label{eq11}
		B_{nm}=(g_{A}/g_{V})^{2} B(GT)_{nm} + B(F)_{nm}.
	\end{equation}
	The constant $D$ value was chosen as 6143 s \cite{Har09} and $g_{A}/g_{V}$ was taken as -1.254 \cite{Nak10}. Many calculations of $\beta$-decay half-lives introduce a quenching factor to reproduce measured data (e.g., authors in Ref.\cite{Tow87} used $[(g_A/g_V)_{eff}]^2 = [0.7(g_A/g_V)_{free}]^2 \sim 0.75 $). Coupling of the weak  forces to two nucleons and existing strong correlations within the nucleus were cited as two important factors to justify quenching of the calculated GT strength~\cite{Gys19}. We did not use any explicit quenching factor in our calculation. The previous half-life  calculations~\cite{Sta90,Hir93}, using the same nuclear model, did not use any explicit quenching factor. This was done because the GT force parameters were so parameterized~\cite{Hom96} in order to reproduce the measured half-lives. The reduced Fermi and GT transition probabilities were explicitly determined using
	\begin{equation} \label{eq12}
		B(F)_{nm} = \frac{1}{2J_{n} +1} \langle{m}\parallel\sum\limits_{k}
		\tau_{-}^{k}\parallel {n}\rangle|^{2}
	\end{equation}
	\begin{equation} \label{eq13}
		B(GT)_{nm} = \frac{1}{2J_{n} +1} \langle{m}\parallel\sum\limits_{k}
		\tau_{-}^{k}\overrightarrow{\sigma}^{k}\parallel {n}\rangle|^{2},
	\end{equation}
	where $\overrightarrow{\sigma}(k)$ and $\tau_{-}^{k}$ denote the spin and the isospin lowering operators, respectively.  For further details on solution of Equation~\ref{eq1} we refer to ~\cite{Hir91,Mut92,Hir93}. The phase space  integrals ($f_{nm}$) over total energy was calculated using  
	\begin{equation} \label{eq14}
		f_{nm} = \int _{1}^{w_{m}}w\sqrt{w^{2} -1}(w_{m} -w)^{2} F(+
		Z, w) (1-R_{-})dw,
	\end{equation}
	where we used natural units ($\hbar=m_{e}=c=1$). The Fermi functions, $F (+Z, w)$, were estimated as per the prescription given in Ref. \cite{Gov71}. 
	$w_{m}$ is the total $\beta$-decay energy given by
	\begin{equation} \label{eq15}
		w_{m} = m_{p} -m_{d} + E_{n} -E_{m}~~~,
	\end{equation}
	\noindent where $E_{n}$ and $E_{m}$, represent parent and daughter excitation energies, respectively. 
	$R_{-}$ is the electron distribution function 
	\begin{equation} \label{eq16}
		R_{-} =\left[\exp \left(\frac{E-E_{f} }{kT} \right)+1\right]^{-1},
	\end{equation}
	where $E$ = ($w$ - 1),  $E_{f}$   denote the kinetic  and Fermi energy of the electrons, respectively.  $k$ is the Boltzmann constant.  As the stellar core temperature rises,  there is always a finite chance of occupation of parent excited levels. The total $\beta$ decay rates were calculated using
	\begin{equation} \label{eq17}
		\lambda^{\beta} =\sum _{nm}P_{n} \lambda _{nm}^{\beta}~~~,
	\end{equation}
	where $P_n$ is the occupation probability of parent excited state following the normal Boltzmann distribution. In Equation~\ref{eq17}, the summation was applied on all final and initial states until reasonable convergence in $\beta$-decay rates was obtained. 
	
	%%%%%%%%%%%%%%%%%%%%%%%%%%%%%%%%%%%%%%%%%%
	\section*{Results and Discussion}
	
	The aim of the current study is to re-examine the effect of pairing gaps on charge-changing transitions and associated weak rates of 50 top-ranked nuclei bearing astrophysical significance and unstable to $\beta^-$ decay~\cite{Aud21}. The nuclei were selected from a recent study by Nabi et al. \cite{Nab21} where a total of  728 nuclei were ranked on the basis of ranking parameter, $\mathring{R}_{p}$, defined by
	\begin{equation} \label{eq19}
		\mathring{R}_{p} = \left(\frac{\dot{Y}^{ec(bd)}_{e(i)}}{\sum\dot{Y}^{ec(bd)}_{e(i)}}\right)_{0.500 > {Y_e} > 0.400},
	\end{equation}
	such that the nuclei having highest $\mathring{R}_p$ value will contribute most to the time-rate of change of lepton fraction ($\dot{Y_e}$). As discussed earlier, three different sets of empirically calculated pairing gaps were used in our analysis to investigate the $\beta$-decay properties of these nuclei. 
	
	The pairing gaps arise from the pairing interaction between nucleons. They have a direct impact on the occupation probabilities of different single-particle states in the nucleus. These probabilities bear consequences for the charge-changing transitions. In general, a larger pairing gap leads to smaller number of nucleons occupying states near the Fermi level. This can contribute to lowering the chances for transitions and may result in redistribution of GT strength to higher excitation energies.  
	
\begin{figure}[!h]
			\centering
		\hspace{-3.0em}\includegraphics[width=6.5in]{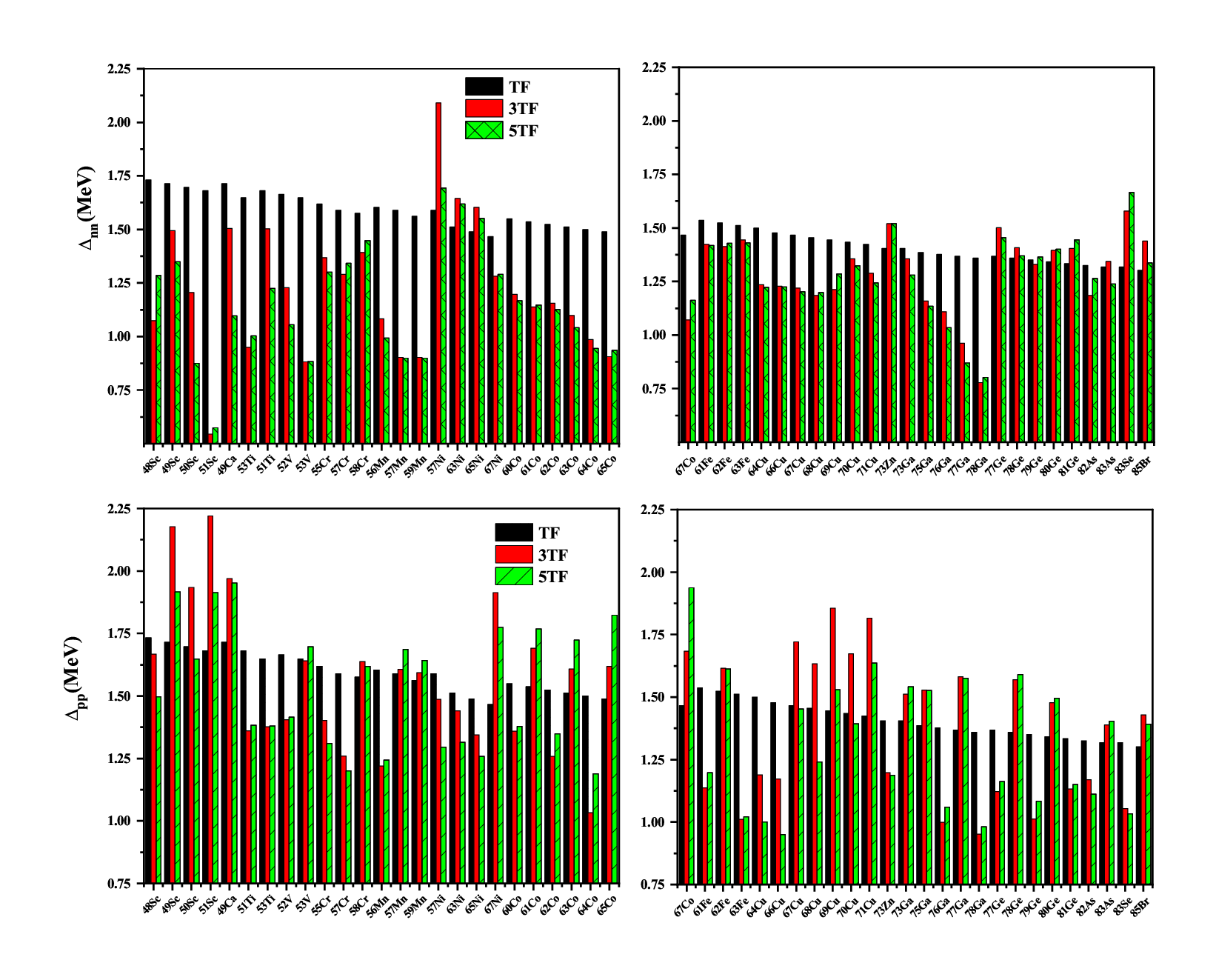}
		\caption{Computed pairing gap values of the selected 50 nuclei used in the current investigation. \label{fig1}}
	\end{figure}

	We first display the computed pairing gaps in Fig. \ref{fig1} for the selected 50 nuclei. The upper panels show the neutron-neutron pairing gaps.   The proton-proton pairing gaps are displayed in the lower panels. The TF formula (Eq.~\ref{eq3}) is only a function of the mass number of the parent nucleus. Nuclear properties of parent and neighbouring nuclei are considered in 3TF formulae (Eq.~\ref{eq4} and Eq.~\ref{eq5}). In  the 5TF formulae (Eq.~\ref{eq6} and Eq.~\ref{eq7}), nuclear properties of two nearest neighbouring nuclei are considered. Table~\ref{Tab1} shows the experimental errors associated with the measured binding energies, used for the computation of 3TF and 5TF schemes. Difference of more than 0.5 MeV in $\Delta_{pp}$ values is noted, between the TF and 3TF schemes, for  
	$^{51}$Sc and $^{63}$Fe. A difference of similar magnitude is noted for $\Delta_{pp}$ between TF and 5TF schemes for the case of $^{64,66}$Cu. The differences between $\Delta_{nn}$ values exceed even more reaching to 0.7 MeV for $^{56}$Mn and more than 1 MeV for $^{51}$Sc.
	%Therefore,  Larger pairing gaps result in an overall decrease in GT strength, while smaller pairing gaps tend to enhance the GT strength within the Q-window. 
	
\begin{figure}[!h]
			\centering
		\hspace{-3.0em}\includegraphics[width=6.0in]{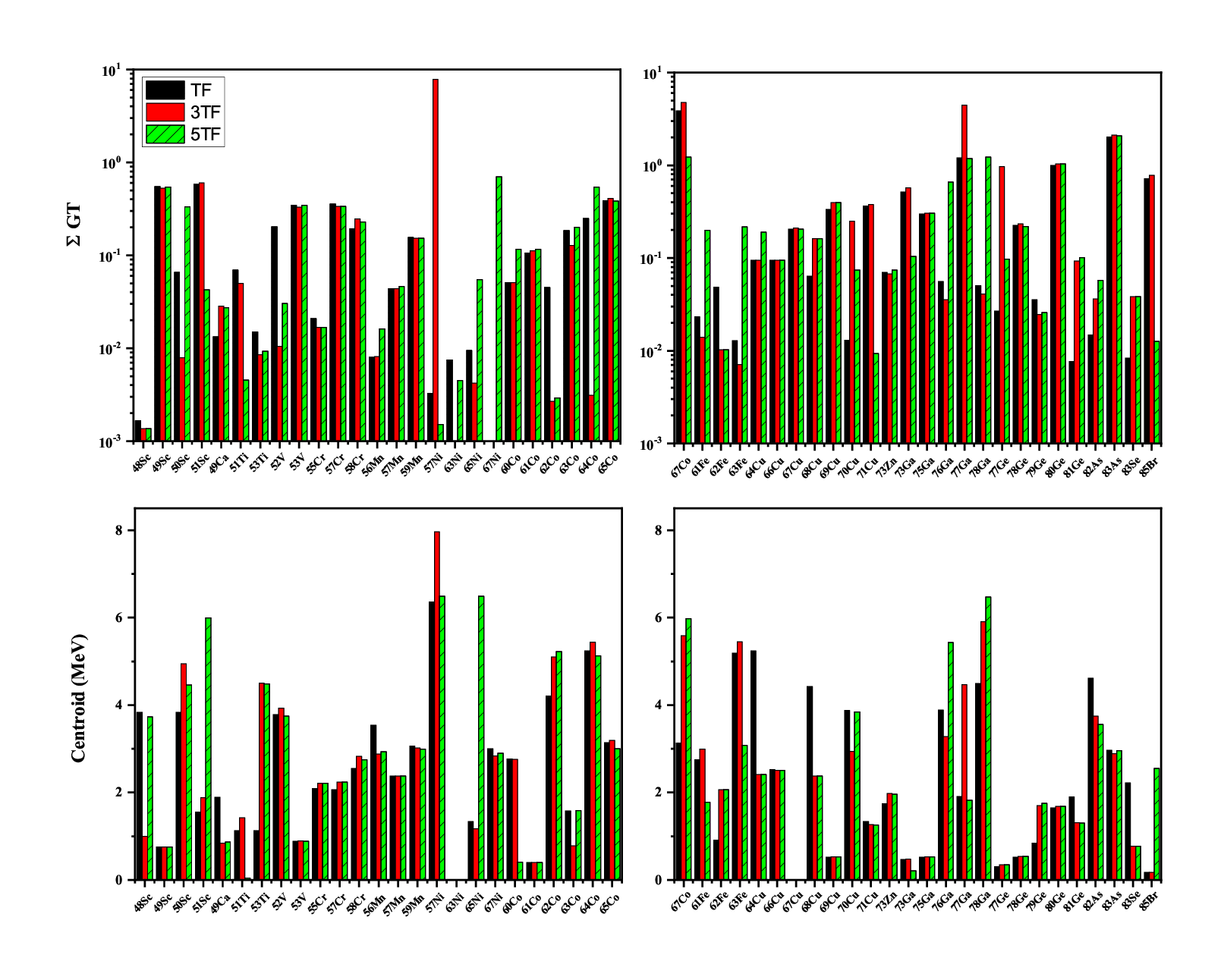}
		\caption{The total strength and centroid values of the calculated GT distributions of the selected 50 nuclei.}\label{fig2}
	\end{figure}

	The total strength and centroid values of the calculated GT strength distributions are shown in Fig.~\ref{fig2} as a function of pairing gap values. The upper panels show the calculated total GT strength whereas the bottom panels show the computed centroids of the resulting distributions. Our calculation satisfied the model-independent Ikeda sum rule~\cite{Ike64}. It is noted from  Fig.~\ref{fig2} that the total strength and centroid values are sensitive function of the pairing gaps. Orders of magnitude differences are noted for the total GT strength as pairing gap value changes. The effect is more pronounced when the $N$ or $Z$ of the nucleus is a magic number. This includes the nuclei $^{57,63,65,67}$Ni, $^{85}$Br. This was expected as changing pairing gap values  would create a bigger impact for closed shell nuclei. It may also be noted that for the case of $^{63,67}$Ni (3TF) and $^{67}$Ni (TF), the total GT strength are smaller than 10$^{-3}$ and therefore not shown in Fig.~\ref{fig2}. The average total GT strength calculated by TF, 3TF and 5TF schemes are 0.30, 0.56 and 0.28, respectively. It was concluded that, overall, the 3TF scheme calculated the largest strength values.  The placement of centroids changes by an order of magnitude or more as we switch from TF to 3TF schemes. The 5TF has a tendency to move the centroid to higher excitation energies whereas the 3TF places the centroid at much lower energies. The average of all centroids computed by TF, 3TF and 5TF are 2.44 MeV, 2.47 MeV and 2.62 MeV, respectively.  More than an order of magnitude difference in the placement of centroid is noted for the case of $^{51}$Ti and $^{85}$Br  (bottom panels of  Fig.~\ref{fig2}). For the case of $^{51}$Ti, only one GT transition was calculated by TF and 3TF schemes at energies 1.1 MeV and 1.4 MeV, respectively. The 5TF schemes calculated more fragmentation of the total strength and at low energies ($<$ 0.1 MeV). This explains the placement of centroids at much higher energies for $^{51}$Ti employing the pairing gap parameter from TF and 3TF schemes.  For the case of $^{85}$Br, the 5TF scheme resulted in high-lying GT transitions (between (2-3) MeV). On the other hand, the TF scheme calculated one GT transition at 2.7 Mev, albeit of magnitude 0.00007. All remaining transitions were within 0.5 MeV in daughter states. The 3TF scheme also computed GT transitions within 0.5 MeV in daughter. Consequently, both TF and 3TF placed the centroid at 0.17 MeV in daughter. 
	
\begin{figure}[!h]
			\centering
		\hspace{-3.0em}\includegraphics[width=6.0in]{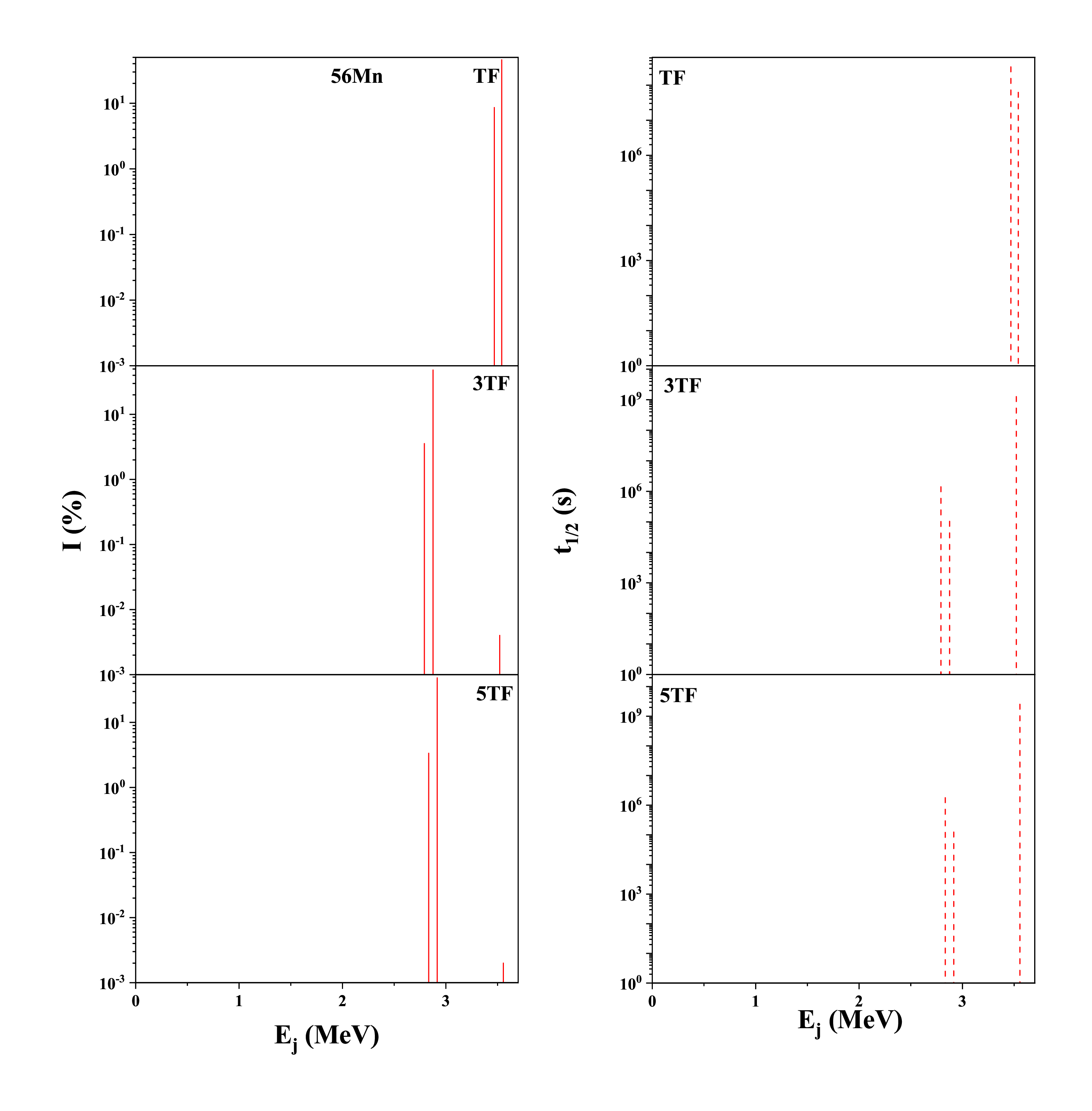}
		\caption{Calculated branching ratios (I) and partial half lives (t$_{1/2}$) for $\beta$ decay of $^{56}$Mn as a function of pairing gaps within the Q-value window. $E_j$ shows excited energy in daughter nucleus.}\label{fig3}
	\end{figure}

\begin{figure}[!h] 
			\centering
		\hspace{-3.0em}\includegraphics[width=6.0in]{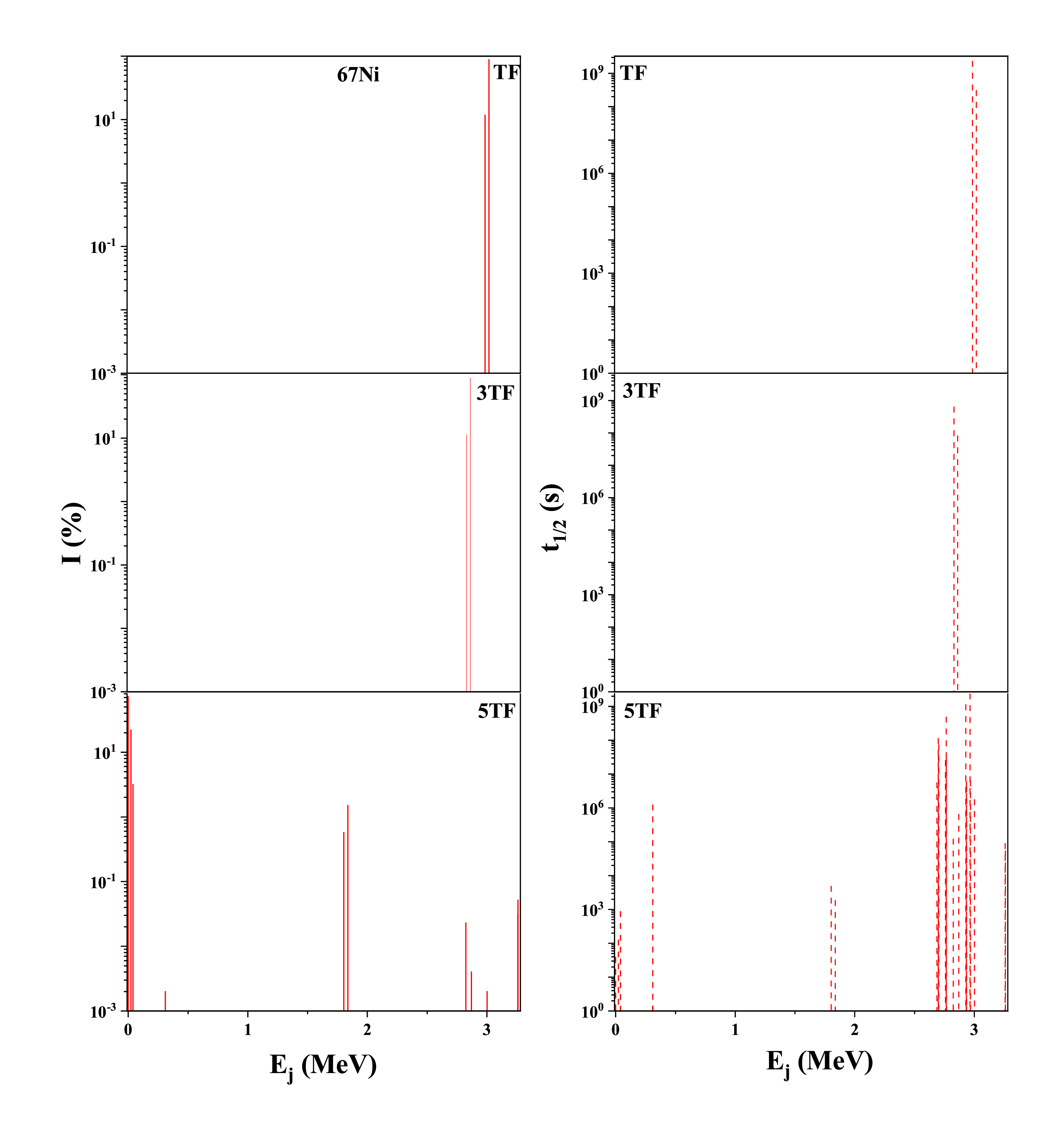}
		\caption{Calculated branching ratios (I) and partial half lives (t$_{1/2}$) for $\beta$ decay of $^{67}$Ni as a function of pairing gaps within the Q-value window. $E_j$ shows excited energy in daughter nucleus.}\label{fig4}
	\end{figure}

\begin{figure}[!h] 
			\centering
		\hspace{-3.0em}\includegraphics[width=6.0in]{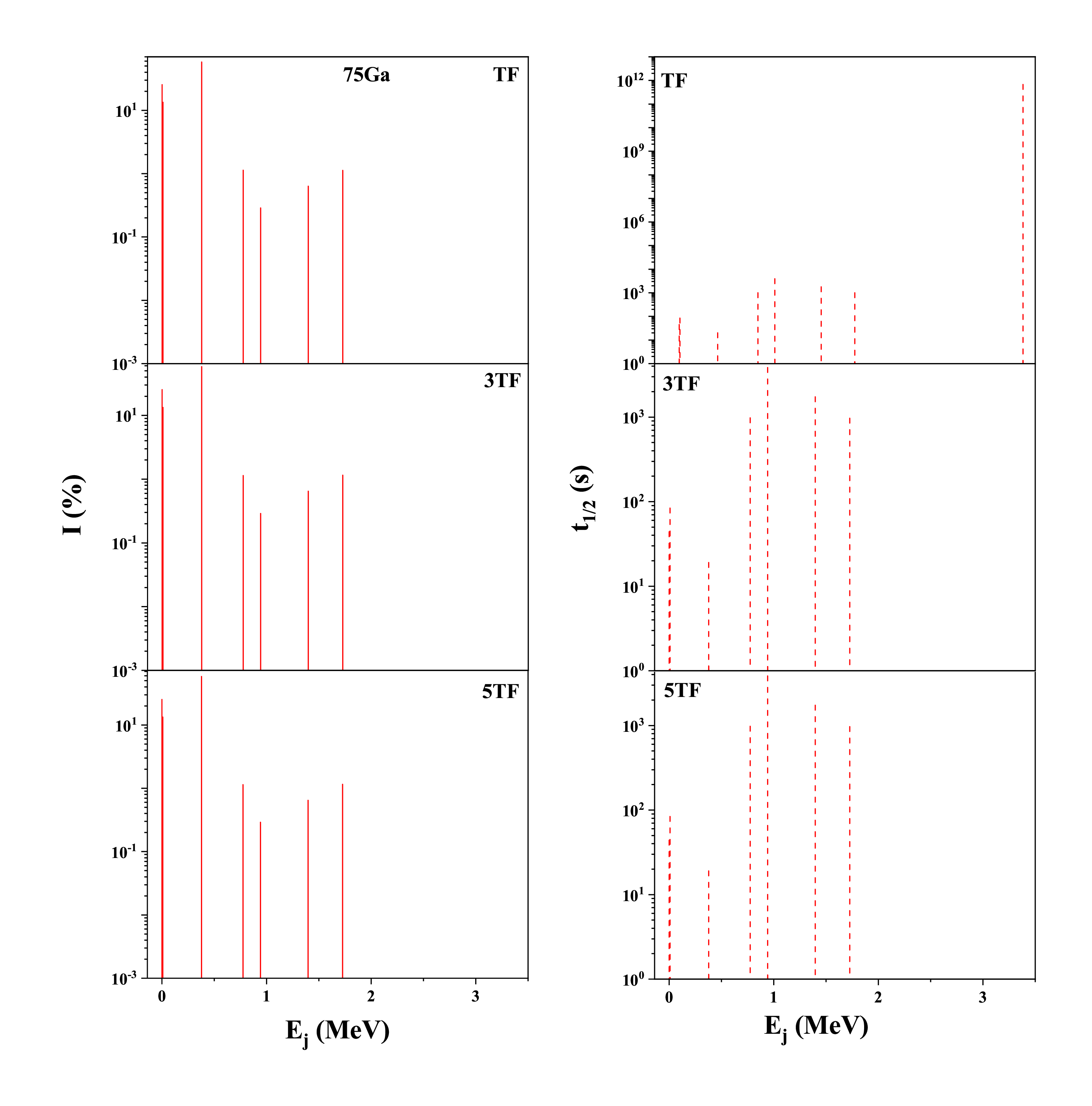}
		\caption{Calculated branching ratios (I) and partial half lives (t$_{1/2}$) for $\beta$ decay of $^{75}$Ga as a function of pairing gaps within the Q-value window. $E_j$ shows excited energy in daughter nucleus.}\label{fig5}
	\end{figure}

\begin{figure}[!h] 
			\centering
		\hspace{-3.0em}\includegraphics[width=6.0in]{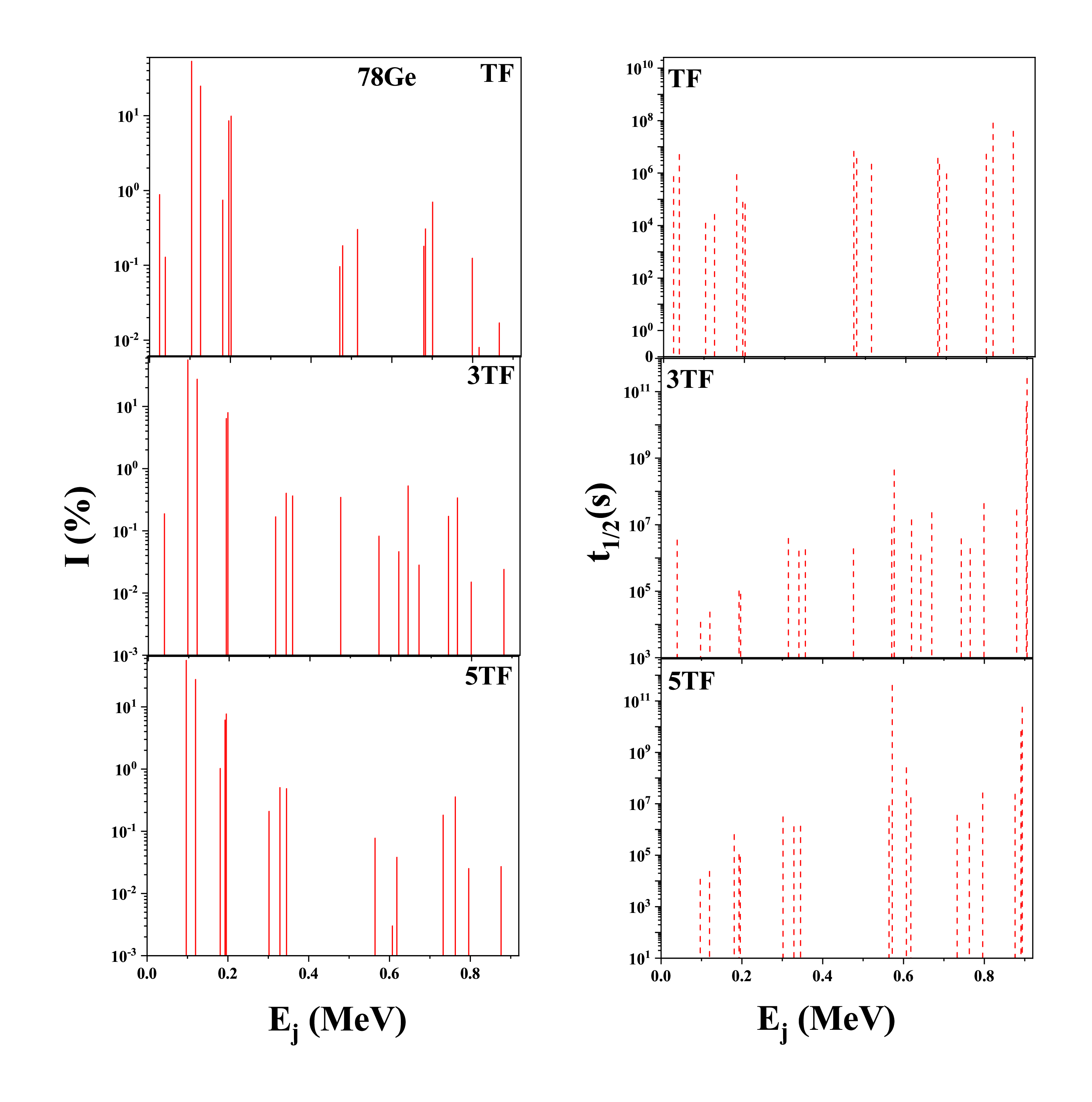}
		\caption{Calculated branching ratios (I) and partial half lives (t$_{1/2}$) for $\beta$ decay of $^{78}$Ge as a function of pairing gaps within the Q-value window. $E_j$ shows excited energy in daughter nucleus.}\label{fig6}
	\end{figure}

	Branching ratios (I) of charge-changing transitions in daughter  was calculated using the equation
	\begin{equation} \label{eq62}
		I = \frac{T_{1/2}}{t_{(1/2)}} \times 100 ~(\%).
	\end{equation}
	Figs.~(\ref{fig3}~-~\ref{fig6}) show the computed branching ratios and partial half-lives as a function of daughter excitation energy for the three selected pairing gaps (TF, 3TF and 5TF) for $^{56}$Mn,  $^{67}$Ni, $^{75}$Ga, and $^{78}$Ge respectively. These nuclei were selected belonging to odd-odd,  even-odd, odd-even and even-even categories from the top-ranked 50 nuclei for the analysis of branching ratios and partial half-lives. Fragmentation of the total GT strength (Fig.~\ref{fig2}) to low-lying states is altered by changing pairing gap values. The effect is different for different classes of nuclei. For odd-odd cases, Fig.~\ref{fig3} shows that low-lying transitions with more fragmentation are produced with 3TF and 5TF schemes. For the magic number nucleus $^{67}$Ni, Fig.~\ref{fig4} shows that the 5TF scheme result in considerable enhancement of the fragmentation of the GT strength when compared with the other two schemes. The 5TF scheme resulted in low lying transition also for the odd-even nucleus $^{75}$Ga as exhibited in  Fig.~\ref{fig5}. For even-even nucleus $^{78}$Ge,  Fig.~\ref{fig6} reveals that the 5TF scheme resulted in lesser fragmentation and high-lying transitions when compared with the TF and 3TF schemes. Equation~\ref*{eq62} helps one to explain $T_{1/2}$ through $t_{1/2}$. The bottom panel of Fig.~\ref{fig3} shows that the $t_{1/2}$ of the decay feeding the state with higher energy is comparable to the other two half-lives, but its branching ratio is almost a hundred times smaller than the others. Likewise, in bottom panel of Fig.~\ref{fig4}, the state at energy 0.31 MeV has a very small branching ratio of 0.002 and hence the contribution of partial half-life is negligible to the total computed half-life.    
	
\begin{figure}[!h]
			\centering
		\hspace{-3.0em}\includegraphics[width=5in]{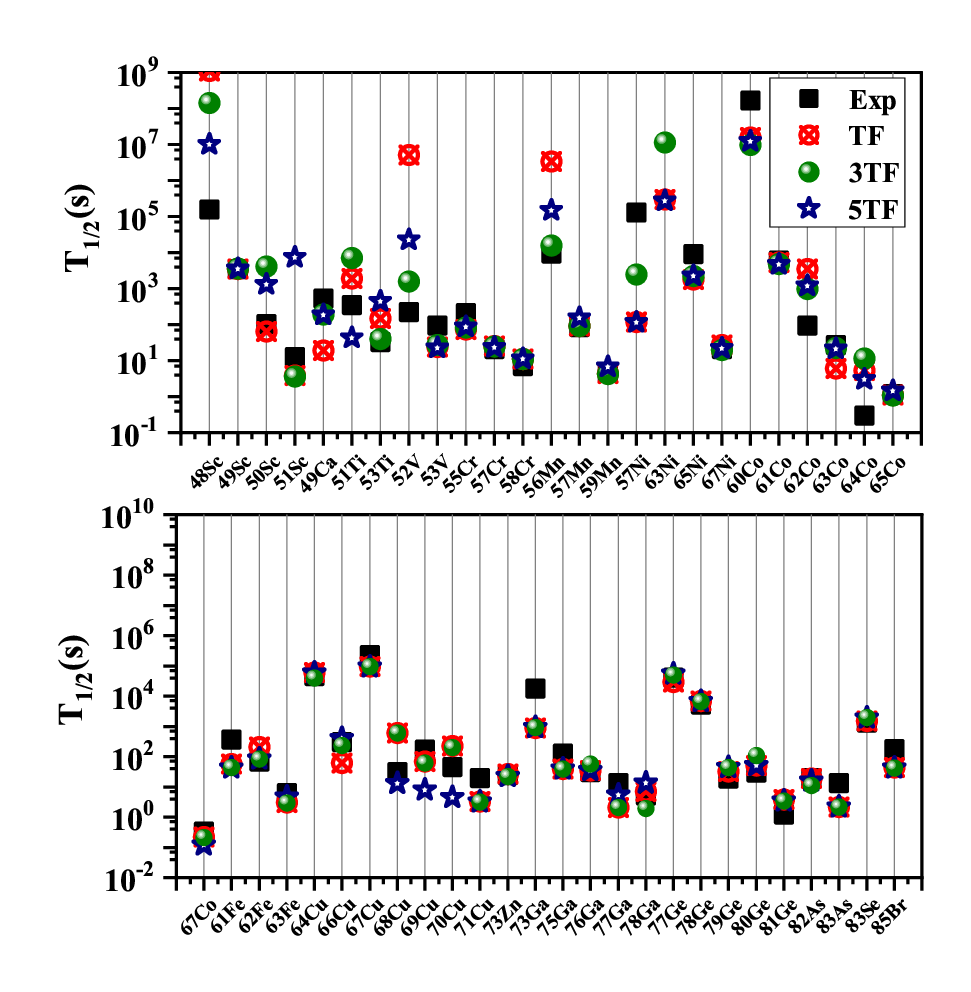}
		\caption{Comparison of measured and predicted half-lives using three different pairing gap values for the selected nuclei. Measured half-lives were taken from Ref.~\cite{Aud21}. \label{fig7}}
	\end{figure}

	The comparison of calculated and measured half-lives for selected top-ranked 50 nuclei is presented in Fig.~\ref{fig7}. The terrestrial  half-lives were calculated using the pn-QRPA model with TF, 3TF and 5TF pairing gap values.  The calculated half-life depends on the total strength and distribution of the GT transitions in the daughter states. These two quantities  were shown earlier in Fig.~\ref{fig2} as a function of the pairing gaps. Three orders of magnitude or more differences in calculated half-life values may be noted from Fig.\ref{fig7}. Higher total GT strength values and lower placement of GT centroid result in smaller calculated half-lives. Table~\ref{Tab2} shows the accuracy of the current nuclear model using different pairing gap values as one of the input parameters. We defined the ratios of calculated to measured half-lives using the variable $y_{i}$  
	\begin{eqnarray}
		y_i = \left\lbrace \frac{T^{cal}_{1/2}}{T^{exp}_{1/2}}~~~~~~ if ~~~~~ T^{cal}_{1/2}\geq T^{exp}_{1/2} \right\rbrace \nonumber\\
		~~~	OR \nonumber\\
		~~~	=\left\lbrace \frac{T^{exp}_{1/2}}{T^{cal}_{1/2}}~~~~~~ if ~~~~~ T^{cal}_{1/2} < T^{exp}_{1/2} \right\rbrace.
	\end{eqnarray} 
	In Table~\ref{Tab2}, $n$ is the number of half-lives (out of a total of 50 cases) reproduced under the condition given in the first column.  The average deviation  ($\overline{y}$) was calculated using
	\begin{equation} \label{eq21}
		y = \frac{1}{n} \sum_{i=1}^n y_i,
	\end{equation} \label{Y}
	Table~\ref{Tab2} shows that the current model with 3TF pairing gap reproduces $80\%$ ($44\%$) of measured half-lives values within a factor of 10 (2) with an average deviation of 2.42 (1.22). We conclude that the 3TF pairing gap results in calculation of bigger total GT strength and best prediction of half-live values for these top-ranked 50 nuclei.
	
	Because of the crucial importance of these nuclei in stellar environment, we decided to calculate $\beta$-decay rates of the selected 50 nuclei as a function of pairing gaps in stellar matter. For $r$-process nuclei and for prevailing physical conditions in stellar matter, forbidden transitions may also contribute to the total weak rates. In the current model we have only calculated allowed GT and Fermi transitions. The calculation of weak rates including forbidden transitions would be taken up as a future assignment.  In general, larger pairing gaps tend to shift the GT centroid to higher excitation energies in daughter. This in turn  decreases the  $\beta$-decay rates.  A larger pairing gap leads to smaller number of nucleons occupying states near the Fermi level. This may result in  redistribution of GT strength to higher excitation energies.  Tables~(\ref{Tab3}-\ref{Tab7}) show the $\beta$ decay rates of top-ranked 50 nuclei at selected densities [$\rho Y_e$ = (10, 10$^5$,  $10^{11}$) g cm$^{-3}$] and temperature [T = (0.1, 1, 5, 10, 15, 30) GK]. In these tables entries written as $<$ 10$^{-100}$ means that the calculated $\beta$-decay rates were less than 10$^{-100}$  $s^{-1}$. Tables~(\ref{Tab3}-\ref{Tab7}) display that $\beta$ decay rates increase as the core temperature increases and decrease as $\rho Y_e$ increases. The decay rates, 
	for a predetermined density,  increase due to accessibility of a large phase space with increasing core temperature. Soaring core temperatures increase the occupation probabilities of
	parent excited levels, thereby leading to a larger contribution of partial rates from parent excited
	states to the total rates. As the stellar core became denser, the electron
	Fermi energy increases leading to a substantial decrease in
	the $\beta$-decay rates at high stellar density values. Specially at high density ($\rho Y_e$ = $10^{11}$ g cm$^{-3}$), because of choking of the phase space, the $\beta$-decay rates tend to zero. It is concluded from Tables~(\ref{Tab3}-\ref{Tab7}) that the 3TF scheme leads to calculation of biggest stellar $\beta$-decay rates. This has a direct correlation with calculation of bigger total GT strength using the 3TF scheme.   Table~\ref{Tab8} shows the average values of the calculated stellar $\beta$-decay rates using different pairing gap values under predetermined physical conditions. 
	
	\section*{Conclusions and Summary}
	In this study, we re-examined the influence of pairing gaps on charge-changing transitions, partial half-lives, branching ratios and weak rates, for top-ranked 50 nuclei having astrophysical significance and unstable to $\beta^-$ decay. Pairing gaps are one of the most important model parameters in the pn-QRPA approach for calculation of $\beta$-decay rates. In order to investigate the effect of pairing gaps on calculated GT strength distributions  and half-lives, we used three different empirically calculated values (referred to as TF, 3TF and 5TF). Changing pairing gap values led to significant alterations in the total GT strength and $\beta$-decay rates. It was concluded that the 3TF pairing gaps resulted in the best prediction of $\beta$-decay half-lives. Following main conclusions are drawn form the current investigation:
	
	$\odot$ The available empirical formulae for pairing gaps give values of $\Delta_{pp}$ differing by 0.5 MeV or more. The difference in  $\Delta_{nn}$ is more than 1 MeV. 
	
	$\odot$ The total GT strength and placement of the computed GT centroid change substantially with the pairing gap values.  The 3TF pairing gap leads to lower placement of GT centroid and higher  total GT strength.
	
	$\odot$ The 3TF scheme gives the best predictive power to the current pn-QRPA model.
	
	$\odot$ The 3TF pairing gaps result in biggest stellar $\beta$-decay rates for the selected top-50 ranked nuclei. 
	
	\begin{table}[htbp]
		\centering
		\caption{The uncertainty associated with the computation of pairing gap values for the selected nuclei.} \label{Tab1}
		\scriptsize
		\renewcommand{\arraystretch}{1.5}
		\renewcommand{\tabcolsep}{0.18cm}
		\renewcommand{\ULdepth}{7.0pt}
		\vspace{-0.1cm}
		\begin{tabular}{ l|l|l|l|l|l|l|l|l }
			
			Nuclei & $\Delta^{3TF}_{pp}$ & $\sigma^{3TF}_{pp}$ & $\Delta^{3TF}_{nn}$ & $\sigma^{3TF}_{nn}$ & $\Delta^{5TF}_{pp}$ & $\sigma^{5TF}_{pp}$ & $\Delta^{5TF}_{nn}$ & $\sigma^{5TF}_{nn}$ \\ \hline
			$^{56}$Mn & 1.2199 & (±0.01262) & 1.0826 & (±0.0110) & 1.2442 & (±0.0001) & 0.9924 & (±0.0002) \\ 
			$^{52}$V & 1.4051 & (±0.0025) & 1.2272 & (±0.0031) & 1.4160 & (±0.0002) & 1.0549 & (±0.0001) \\ 
			$^{67}$Cu & 1.7205 & (±0.0020) & 1.2201 & (±0.0015) & 1.4529 & (±0.0001) & 1.2019 & (±0.0001) \\ 
			$^{67}$Ni & 1.9130 & (±0.0139) & 1.2822 & (±0.0029) & 1.7737 & (±0.0001) & 1.2899 & (±0.0002) \\ 
			$^{60}$Co & 1.3590 & (±0.0021) & 1.1972 & (±0.0011) & 1.3787 & (±0.0002) & 1.1672 & (±0.0001) \\ 
			$^{49}$Sc & 2.1768 & (±0.0022) & 1.4937 & (±0.0049) & 1.9171 & (±0.0001) & 1.3496 & (±0.0001) \\ 
			$^{66}$Cu & 1.1727 & (±0.02000) & 1.2278 & (±0.0008) & 0.9495 & (±0.0001) & 1.2239 & (±0.0003) \\ 
			$^{50}$Sc & 1.9343 & (±0.0025) & 1.2055 & (±0.0049) & 1.6477 & (±0.0001) & 0.8751 & (±0.0001) \\ 
			$^{79}$Ge & 1.0129 & (±0.0371) & 1.3296 & (±0.0371) & 1.0833 & (±0.0004) & 1.3652 & (±0.0004) \\ 
			$^{65}$Co & 1.6184 & (±0.0050) & 0.9056 & (±0.0200) & 1.8231 & (±0.0003) & 0.9350 & (±0.0001) \\ 
			$^{63}$Co & 1.6088 & (±0.0185) & 1.0975 & (±0.0200) & 1.7241 & (±0.0003) & 1.0406 & (±0.0002) \\ 
			$^{77}$Ga & 1.5809 & (±0.0035) & 0.9609 & (±0.0024) & 1.5749 & (±0.0001) & 0.8700 & (±0.0001) \\ 
			$^{78}$Ge & 1.5680 & (±0.0053) & 1.4075 & (±0.0371) & 1.5901 & (±0.0004) & 1.3700 & (±0.0001) \\ 
			$^{83}$As & 1.3892 & (±0.0032) & 1.3430 & (±0.0037) & 1.4039 & (±0.0001) & 1.2378 & (±0.0001) \\ 
			$^{51}$Ti & 1.3611 & (±0.0025) & 1.5025 & (±0.0027) & 1.3831 & (±0.0001) & 1.2243 & (±0.0001) \\ 
			$^{59}$Mn & 1.5931 & (±0.0847) & 0.9028 & (±0.0027) & 1.6418 & (±0.0001) & 0.8997 & (±0.0014) \\ 
			$^{64}$Co & 1.0334 & (±0.0200) & 0.9853 & (±0.0200) & 1.1886 & (±0.0003) & 0.9449 & (±0.0003) \\ 
			$^{49}$Ca & 1.9691 & (±0.0025) & 1.5047 & (±0.0022) & 1.9517 & (±0.0001) & 1.0974 & (±0.0001) \\ 
			$^{58}$Cr & 1.6375 & (±0.1002) & 1.3923 & (±0.0029) & 1.6181 & (±0.0001) & 1.4475 & (±0.0017) \\ 
			$^{68}$Cu & 1.6340 & (±0.0139) & 1.1839 & (±0.0015) & 1.2401 & (±0.0001) & 1.1982 & (±0.0002) \\ 
			$^{82}$As & 1.1693 & (±0.0037) & 1.1848 & (±0.0037) & 1.1112 & (±0.0001) & 1.2637 & (±0.0001) \\ 
			$^{75}$Ga & 1.5284 & (±0.0025) & 1.1589 & (±0.0029) & 1.5263 & (±0.0001) & 1.1338 & (±0.0001) \\ 
			$^{69}$Cu & 1.8563 & (±0.0064) & 1.2126 & (±0.0015) & 1.5295 & (±0.0001) & 1.2842 & (±0.0001) \\ 
			$^{57}$Ni & 1.4860 & (±0.0005) & 2.0909 & (±0.0007) & 1.2959 & (±0.0001) & 1.6936 & (±0.0001) \\ 
			$^{61}$Fe & 1.1374 & (±0.0185) & 1.4226 & (±0.0034) & 1.1978 & (±0.0001) & 1.4176 & (±0.0002) \\ 
			$^{81}$Ge & 1.1328 & (±0.0037) & 1.4039 & (±0.0371) & 1.1508 & (±0.0004) & 1.4434 & (±0.0001) \\ 
			$^{78}$Ga & 0.9509 & (±0.0371) & 0.7791 & (±0.0024) & 0.9806 & (±0.0001) & 0.8022 & (±0.0004) \\ 
			$^{51}$Sc & 2.2206 & (±0.0027) & 0.5445 & (±0.0030) & 1.9128 & (±0.0003) & 0.5739 & (±0.0001) \\ 
			$^{64}$Cu & 1.1883 & (±0.0185) & 1.2356 & (±0.0006) & 1.0006 & (±0.0001) & 1.2227 & (±0.0002) \\ 
			$^{57}$Cr & 1.2598 & (±0.1758) & 1.2904 & (±0.0029) & 1.2004 & (±0.0001) & 1.3414 & (±0.0031) \\ 
			$^{77}$Ge & 1.1226 & (±0.0097) & 1.5012 & (±0.0035) & 1.1623 & (±0.0004) & 1.4548 & (±0.0001) \\ 
			$^{55}$Cr & 1.4016 & (±0.0111) & 1.3681 & (±0.0005) & 1.3107 & (±0.0001) & 1.3009 & (±0.0002) \\ 
			$^{83}$Se & 1.0541 & (±0.0257) & 1.5793 & (±0.0030) & 1.0326 & (±0.0001) & 1.6650 & (±0.0003) \\ 
			$^{62}$Fe & 1.6161 & (±0.0185) & 1.4124 & (±0.0043) & 1.6126 & (±0.0001) & 1.4283 & (±0.0002) \\ 
			$^{48}$Sc & 1.6681 & (±0.00495) & 1.0748 & (±0.004) & 1.4966 & (±0.0001) & 1.2842 & (±0.0001) \\ 
			$^{65}$Ni & 1.3447 & (±0.02000) & 1.6032 & (±0.0013) & 1.2584 & (±0.0001) & 1.5513 & (±0.0003) \\ 
			$^{57}$Mn & 1.6066 & (±0.0270) & 0.9021 & (±0.0027) & 1.6861 & (±0.0001) & 0.8998 & (±0.0004) \\ 
			$^{71}$Cu & 1.8159 & (±0.0856) & 1.2893 & (±0.0014) & 1.6357 & (±0.0001) & 1.2441 & (±0.0012) \\ 
			$^{53}$Ti & 1.3766 & (±0.0111) & 0.9499 & (±0.0158) & 1.3804 & (±0.0005) & 1.0040 & (±0.0002) \\ 
			$^{53}$V & 1.6405 & (±0.0031) & 0.8829 & (±0.0118) & 1.6974 & (±0.0004) & 0.8841 & (±0.0001) \\ 
			$^{73}$Ga & 1.5124 & (±0.0021) & 1.3556 & (±0.0029) & 1.5412 & (±0.0001) & 1.2805 & (±0.0001) \\ 
			$^{85}$Br & 1.4286 & (±0.0030) & 1.4394 & (±0.02573) & 1.3917 & (±0.0003) & 1.3371 & (±0.0001) \\ 
			$^{62}$Co & 1.2586 & (±0.0185) & 1.1560 & (±0.0185) & 1.3493 & (±0.0003) & 1.1262 & (±0.0002) \\ 
			$^{70}$Cu & 1.6734 & (±0.0038) & 1.3558 & (±0.0015) & 1.3925 & (±0.0001) & 1.3226 & (±0.0001) \\ 
			$^{76}$Ga & 0.9984 & (±0.0061) & 1.1087 & (±0.0029) & 1.0605 & (±0.0001) & 1.0347 & (±0.0001) \\ 
			$^{73}$Zn & 1.1969 & (±0.0029) & 1.5210 & (±0.0026) & 1.1866 & (±0.0001) & 1.5200 & (±0.0001) \\ 
			$^{80}$Ge & 1.4776 & (±0.00264) & 1.3965 & (±0.0371) & 1.4952 & (±0.0004) & 1.4016 & (±0.0001) \\ 
			$^{63}$Ni & 1.4407 & (±0.0185) & 1.6443 & (±0.0004) & 1.3143 & (±0.0001) & 1.6195 & (±0.0002) \\ 
			$^{67}$Co & 1.6837 & (±0.0064) & 1.0702 & (±0.0139) & 1.9365 & (±0.0012) & 1.1621 & (±0.0001) \\ 
			$^{61}$Co & 1.6913 & (±0.0034) & 1.1372 & (±0.0185) & 1.7680 & (±0.0002) & 1.1463 & (±0.0001) \\ 
			$^{63}$Fe & 1.0101 & (±0.0200) & 1.4439 & (±0.0050) & 1.0213 & (±0.0001) & 1.4297 & (±0.0003) \\ 
		\end{tabular}
	\end{table}
	\clearpage

	
	\begin{table}[htbp]
		\centering
		\caption{Accuracy of the pn-QRPA model calculated half-lives using three different pairing gaps for the selected 50 top-ranked $\beta$-decaying nuclei.} \label{Tab2}
\begin{tabular}{l| l| l l l }
			Condition  &  Pairing Gaps &  n &  n\% &  $\overline{y}$  \\ \hline
			$\forall \hspace*{0.2cm} y_{i}$ $\leq$ 10 & TF & 39 & 78 & 2.81 \\ 
			& 3TF & 40 & 80 & 2.42 \\
			& 5TF & 36 & 72 & 3.18 \\
			\hline
			$\forall \hspace*{0.2cm} y_{i}$ $\leq$ 2  & TF & 21 & 42 & 1.32  \\ 
			& 3TF & 22 & 44 & 1.22 \\
			& 5TF & 18 & 36 & 1.42\\
		\end{tabular}
	\end{table} 
	
	\clearpage
	\begin{table}[htbp]
		\centering
		\caption{Comparison of calculated stellar $\beta$ decay rates for $^{48-51}$Sc, $^{49}$Ca, $^{51,53}$Ti, $^{52,53}$V, $^{55}$Cr and $^{56}$Mn as a function of pairing gap values. The stellar core temperature is given in units of $GK$ and densities in units of $g  cm ^{-3}$.}
		\scriptsize
		\renewcommand{\arraystretch}{1.5}
		\renewcommand{\tabcolsep}{0.04cm}
		\renewcommand{\ULdepth}{7.0pt}
		\vspace{-0.1cm}
		\begin{tabular}{c|c|ccc|ccc|ccc}
			{\textbf{Nuclei}} &\textbf{T} & \multicolumn{3}{c}{\textbf{$\rho$=$10^7$}} & \multicolumn{3}{c}{\textbf{$\rho$=$10^9$}} & \multicolumn{3}{c}{\textbf{$\rho$=$10^{11}$}} 
			\\
			\hline
			&       & \textbf{TF} & \textbf{3TF} & \textbf{5TF} & \textbf{TF} & \textbf{3TF} & \textbf{5TF} & \textbf{TF} & \textbf{3TF} & \textbf{5TF} \\
			\hline
			$^{48}$Sc & 5 & 9.16$\times10^{-05}$ & 1.00$\times10^{-03}$ & 4.51$\times10^{-04}$ & 2.06$\times10^{-07}$ & 1.60$\times10^{-06}$ & 8.75$\times10^{-07}$ & 8.02$\times10^{-26}$ & 1.42$\times10^{-24}$ & 4.13$\times10^{-25}$ \\ 
			~ & 10 & 5.45$\times10^{-03}$ & 1.10$\times10^{-02}$ & 1.01$\times10^{-02}$ & 5.33$\times10^{-04}$ & 1.24$\times10^{-03}$ & 1.24$\times10^{-03}$ & 2.59$\times10^{-13}$ & 1.00$\times10^{-12}$ & 6.55$\times10^{-13}$ \\ 
			~ & 15 & 3.49$\times10^{-02}$ & 7.36$\times10^{-02}$ & 7.18$\times10^{-02}$ & 9.86$\times10^{-03}$ & 2.32$\times10^{-02}$ & 2.25$\times10^{-02}$ & 5.26$\times10^{-09}$ & 1.54$\times10^{-08}$ & 1.24$\times10^{-08}$ \\ 
			~ & 30 & 2.43$\times10^{-01}$ & 6.10$\times10^{-01}$ & 5.73$\times10^{-01}$ & 1.87$\times10^{-01}$ & 4.76$\times10^{-01}$ & 4.46$\times10^{-01}$ & 1.44$\times10^{-04}$ & 3.91$\times10^{-04}$ & 3.50$\times10^{-04}$ \\ 
			$^{49}$Ca & 5 & 5.62$\times10^{-02}$ & 3.46$\times10^{-02}$ & 2.17$\times10^{-02}$ & 4.04$\times10^{-03}$ & 2.54$\times10^{-03}$ & 2.20$\times10^{-03}$ & 5.47$\times10^{-21}$ & 4.04$\times10^{-21}$ & 4.07$\times10^{-21}$ \\ 
			~ & 10 & 3.80$\times10^{-01}$ & 2.52$\times10^{-01}$ & 2.40$\times10^{-01}$ & 1.25$\times10^{-01}$ & 8.45$\times10^{-02}$ & 8.59$\times10^{-02}$ & 1.14$\times10^{-10}$ & 9.12$\times10^{-11}$ & 8.73$\times10^{-11}$ \\
\end{tabular}%
		\label{Tab3}%
	\end{table}%
	
	\begin{table}[htbp]
		\centering
		\caption{Comparison of calculated stellar $\beta$ decay rates for $^{57,58}$Cr, $^{57,59}$Mn, $^{57}$Ni, $^{60-63}$Co and $^{61-63}$Fe as a function of pairing gap values. The stellar core temperature is given in units of $GK$ and densities in units of $g  cm ^{-3}$.}
		\scriptsize
		\renewcommand{\arraystretch}{1.5}
		\renewcommand{\tabcolsep}{0.04cm}
		\renewcommand{\ULdepth}{6.0pt}
		\vspace{0.5cm}
		\begin{tabular}{c|c|ccc|ccc|ccc}
			{\textbf{Nuclei}} &\textbf{T} & \multicolumn{3}{c}{\textbf{$\rho$=$10^7$}} & \multicolumn{3}{c}{\textbf{$\rho$=$10^9$}} & \multicolumn{3}{c}{\textbf{$\rho$=$10^{11}$}} 
			\\
			\hline
			&       & \textbf{TF} & \textbf{3TF} & \textbf{5TF} & \textbf{TF} & \textbf{3TF} & \textbf{5TF} & \textbf{TF} & \textbf{3TF} & \textbf{5TF} \\
			\hline 
			$^{57}$Cr & 5 & 1.31$\times10^{-02}$ & 1.17$\times10^{-01}$ & 1.16$\times10^{-01}$ & 1.68$\times10^{-03}$ & 7.64$\times10^{-03}$ & 7.64$\times10^{-03}$ & 1.46$\times10^{-20}$ & 2.59$\times10^{-21}$ & 2.67$\times10^{-21}$ \\ 
			~ & 10 & 1.34$\times10^{-01}$ & 3.23$\times10^{-01}$ & 3.24$\times10^{-01}$ & 4.75$\times10^{-02}$ & 7.38$\times10^{-02}$ & 7.43$\times10^{-02}$ & 2.04$\times10^{-010}$ & 4.03$\times10^{-11}$ & 4.09$\times10^{-11}$ \\ 
			~ & 15 & 6.84$\times10^{-01}$ & 8.87$\times10^{-01}$ & 8.55$\times10^{-01}$ & 3.69$\times10^{-01}$ & 3.61$\times10^{-01}$ & 3.46$\times10^{-01}$ & 8.39$\times10^{-07}$ & 2.40$\times10^{-07}$ & 2.21$\times10^{-07}$ \\ 
			~ & 30 & 4.63$\times10^{+00}$ & 5.43$\times10^{+00}$ & 3.44$\times10^{+00}$ & 3.94$\times10^{+00}$ & 4.43$\times10^{+00}$ & 2.77$\times10^{+00}$ & 6.05$\times10^{-03}$ & 4.50$\times10^{-03}$ & 2.55$\times10^{-03}$ \\ 
			$^{57}$Mn & 5 & 8.81$\times10^{-03}$ & 8.32$\times10^{-03}$ & 8.32$\times10^{-03}$ & 1.51$\times10^{-05}$ & 1.36$\times10^{-05}$ & 1.36$\times10^{-05}$ & 3.46$\times10^{-24}$ & 2.13$\times10^{-24}$ & 2.66$\times10^{-24}$ \\ 
			~ & 10 & 1.59$\times10^{-02}$ & 1.31$\times10^{-02}$ & 1.31$\times10^{-02}$ & 1.27$\times10^{-03}$ & 7.80$\times10^{-04}$ & 8.09$\times10^{-04}$ & 5.85$\times10^{-13}$ & 3.22$\times10^{-13}$ & 3.72$\times10^{-13}$ \\
\end{tabular}%
		\label{Tab4}%
	\end{table}%
	
	\begin{table}[htbp]
		\centering
		\caption{Comparison of calculated stellar $\beta$ decay rates for $^{63,65,67}$Ni, $^{64,65,67}$Co and $^{64,66-70}$Cu as a function of pairing gap values. The stellar core temperature is given in units of $GK$ and densities in units of $g  cm ^{-3}$.}
		\scriptsize
		\renewcommand{\arraystretch}{1.5}
		\renewcommand{\tabcolsep}{0.01cm}
		\renewcommand{\ULdepth}{6.0pt}
		\vspace{0.8cm}
		\begin{tabular}{c|c|ccc|ccc|ccc}
			{\textbf{Nuclei}} &\textbf{T} & \multicolumn{3}{c}{\textbf{$\rho$=$10^7$}} & \multicolumn{3}{c}{\textbf{$\rho$=$10^9$}} & \multicolumn{3}{c}{\textbf{$\rho$=$10^{11}$}} 
			\\
			\hline
			&       & \textbf{TF} & \textbf{3TF} & \textbf{5TF} & \textbf{TF} & \textbf{3TF} & \textbf{5TF} & \textbf{TF} & \textbf{3TF} & \textbf{5TF} \\
			\hline
			$^{63}$Ni & 5 & 1.02$\times10^{-05}$ & 9.73$\times10^{-06}$ & 8.69$\times10^{-06}$ & 6.25$\times10^{-09}$ & 7.48$\times10^{-09}$ & 7.06$\times10^{-09}$ & 6.59$\times10^{-28}$ & 8.83$\times10^{-28}$ & 7.38$\times10^{-28}$ \\ 
			~ & 10 & 3.30$\times10^{-04}$ & 3.91$\times10^{-04}$ & 2.35$\times10^{-04}$ & 1.82$\times10^{-05}$ & 3.95$\times10^{-05}$ & 1.31$\times10^{-05}$ & 4.86$\times10^{-15}$ & 6.46$\times10^{-14}$ & 3.50$\times10^{-15}$ \\ 
			~ & 15 & 3.26$\times10^{-03}$ & 1.01$\times10^{-02}$ & 1.97$\times10^{-03}$ & 6.89$\times10^{-04}$ & 4.23$\times10^{-03}$ & 4.10$\times10^{-04}$ & 2.45$\times10^{-10}$ & 7.19$\times10^{-09}$ & 1.45$\times10^{-10}$ \\ 
			~ & 30 & 3.29$\times10^{-02}$ & 7.46$\times10^{-01}$ & 1.85$\times10^{-02}$ & 2.43$\times10^{-02}$ & 6.32$\times10^{-01}$ & 1.37$\times10^{-02}$ & 1.52$\times10^{-05}$ & 9.77$\times10^{-04}$ & 8.43$\times10^{-06}$ \\ 
			$^{64}$Co & 5 & 5.22$\times10^{-01}$ & 1.02$\times10^{+00}$ & 9.79$\times10^{-01}$ & 1.03$\times10^{-02}$ & 6.97$\times10^{-02}$ & 6.65$\times10^{-02}$ & 1.48$\times10^{-21}$ & 1.50$\times10^{-20}$ & 1.50$\times10^{-20}$ \\ 
			~ & 10 & 8.17E-01 & 2.00$\times10^{+00}$ & 1.86$\times10^{+00}$ & 1.06$\times10^{-01}$ & 3.94$\times10^{-01}$ & 3.66$\times10^{-01}$ & 3.89$\times10^{-11}$ & 1.67$\times10^{-10}$ & 1.57$\times10^{-10}$ \\
\end{tabular}%
		\label{Tab5}%
	\end{table}%
	
	\begin{table}[htbp]
		\centering
		\caption{Comparison of calculated stellar $\beta$ decay rates for $^{71}$Cu, $^{73,75-78}$Ga, $^{73}$Zn and $^{77-80}$Ge as a function of pairing gap values. The stellar core temperature is given in units of $GK$ and densities in units of $g  cm ^{-3}$.}
		\scriptsize
		\renewcommand{\arraystretch}{1.5}
		\renewcommand{\tabcolsep}{0.04cm}
		\renewcommand{\ULdepth}{6.0pt}
		\vspace{0.5cm}
		\begin{tabular}{c|c|ccc|ccc|ccc}
			{\textbf{Nuclei}} &\textbf{T} & \multicolumn{3}{c}{\textbf{$\rho$=$10^7$}} & \multicolumn{3}{c}{\textbf{$\rho$=$10^9$}} & \multicolumn{3}{c}{\textbf{$\rho$=$10^{11}$}} 
			\\
			\hline
			&       & \textbf{TF} & \textbf{3TF} & \textbf{5TF} & \textbf{TF} & \textbf{3TF} & \textbf{5TF} & \textbf{TF} & \textbf{3TF} & \textbf{5TF} \\
			\hline
			$^{71}$Cu & 5 & 3.85$\times10^{-01}$ & 3.45$\times10^{-01}$ & 3.52$\times10^{-01}$ & 6.95$\times10^{-01}$ & 5.51$\times10^{-03}$ & 5.69$\times10^{-03}$ & 1.18$\times10^{-21}$ & 1.05$\times10^{-21}$ & 1.08$\times10^{-21}$ \\ 
			~ & 10 & 6.17$\times10^{-01}$ & 5.50$\times10^{-01}$ & 5.77$\times10^{-01}$ & 7.78$\times10^{-02}$ & 6.62$\times10^{-02}$ & 7.00$\times10^{-02}$ & 3.00$\times10^{-11}$ & 2.90$\times10^{-11}$ & 2.83$\times10^{-11}$ \\ 
			~ & 15 & 1.30E$\times10^{+00}$ & 1.12$\times10^{+00}$ & 1.17$\times10^{+00}$ & 4.00$\times10^{-01}$ & 3.41$\times10^{-01}$ & 3.55$\times10^{-01}$ & 1.95$\times10^{-07}$ & 1.96$\times10^{-07}$ & 1.77$\times10^{-07}$ \\ 
			~ & 30 & 4.19$\times10^{+00}$ & 4.29$\times10^{+00}$ & 3.53$\times10^{+00}$ & 3.27$\times10^{+00}$ & 3.36$\times10^{+00}$ & 2.74$\times10^{+00}$ & 2.54$\times10^{-03}$ & 2.99$\times10^{-03}$ & 2.13$\times10^{-03}$ \\ 
			$^{73}$Ga & 5 & 3.56$\times10^{-03}$ & 3.67$\times10^{-03}$ & 3.69$\times10^{-03}$ & 7.96$\times10^{-06}$ & 7.48$\times10^{-06}$ & 7.26$\times10^{-06}$ & 1.05$\times10^{-24}$ & 1.04$\times10^{-24}$ & 9.55$\times10^{-25}$ \\ 
			~ & 10 & 2.74$\times10^{-03}$ & 2.56$\times10^{-03}$ & 2.72$\times10^{-03}$ & 2.01$\times10^{-03}$ & 1.82$\times10^{-03}$ & 1.87$\times10^{-03}$ & 6.37$\times10^{-13}$ & 6.53$\times10^{-13}$ & 5.92$\times10^{-13}$ \\
\end{tabular}%
		\label{Tab6}%
	\end{table}%
	
	\begin{table}[htbp]
		\centering
		\caption{Comparison of calculated stellar $\beta$ decay rates for $^{81}$Ge, $^{82,83}$As, $^{83}$Se and $^{85}$Br as a function of pairing gap values. The stellar core temperature is given in units of $GK$ and densities in units of $g  cm ^{-3}$.}
		\scriptsize
		\renewcommand{\arraystretch}{1.5}
		\renewcommand{\tabcolsep}{0.04cm}
		\renewcommand{\ULdepth}{6.0pt}
		\vspace{0.5cm}
		\begin{tabular}{c|c|ccc|ccc|ccc}
			{\textbf{Nuclei}} &\textbf{T} & \multicolumn{3}{c}{\textbf{$\rho$=$10^7$}} & \multicolumn{3}{c}{\textbf{$\rho$=$10^9$}} & \multicolumn{3}{c}{\textbf{$\rho$=$10^{11}$}} 
			\\
			\hline
			&       & \textbf{TF} & \textbf{3TF} & \textbf{5TF} & \textbf{TF} & \textbf{3TF} & \textbf{5TF} & \textbf{TF} & \textbf{3TF} & \textbf{5TF} \\
			\hline
			$^{81}$Ge & 5 & 2.90$\times10^{-02}$ & 2.94$\times10^{-02}$ & 3.04$\times10^{-02}$ & 2.62$\times10^{-03}$ & 2.67$\times10^{-03}$ & 2.86$\times10^{-03}$ & 1.77$\times10^{-20}$ & 1.45$\times10^{-20}$ & 1.52$\times10^{-20}$ \\ 
			~ & 10 & 4.17$\times10^{-01}$ & 4.12$\times10^{-01}$ & 4.12$\times10^{-01}$ & 1.50$\times10^{-01}$ & 1.43$\times10^{-01}$ & 1.43$\times10^{-01}$ & 2.42$\times10^{-10}$ & 2.16$\times10^{-10}$ & 2.05$\times10^{-10}$ \\ 
			~ & 15 & 3.05$\times10^{+00}$ & 3.06$\times10^{+00}$ & 2.84$\times10^{+00}$ & 1.58$\times10^{+00}$ & 1.57$\times10^{+00}$ & 1.45$\times10^{+00}$ & 1.85$\times10^{-06}$ & 1.84$\times10^{-06}$ & 1.59$\times10^{-06}$ \\ 
			~ & 30 & 2.64$\times10^{+01}$ & 3.05$\times10^{+01}$ & 2.37$\times10^{+01}$ & 2.21$\times10^{+01}$ & 2.56$\times10^{+01}$ & 1.98$\times10^{+01}$ & 2.68$\times10^{-02}$ & 3.15$\times10^{-02}$& 2.33$\times10^{-02}$ \\ 
			$^{82}$As & 5 & 3.46$\times10^{-02}$ & 4.58$\times10^{-02}$ & 4.20$\times10^{-02}$ & 1.16$\times10^{-03}$ & 1.63$\times10^{-03}$ & 1.61$\times10^{-03}$ & 1.08$\times10^{-21}$ & 2.13$\times10^{-21}$ & 2.16$\times10^{-21}$ \\ 
			~ & 10 & 2.97$\times10^{-01}$ & 3.93$\times10^{-01}$ & 3.79$\times10^{-01}$ & 7.16$\times10^{-02}$ & 1.06$\times10^{-01}$ & 1.04$\times10^{-01}$ & 5.31$\times10^{-11}$ & 8.87$\times10^{-11}$ & 8.77$\times10^{-11}$ \\
\end{tabular}%
		\label{Tab7}%
	\end{table}%
	\begin{table}[htbp]
		\centering
		\caption{Average $\beta$-decay rates calculated using different pairing gap values for limiting physical conditions stated in the heading.  The stellar core temperature is given in units of $GK$ and densities in units of $g  cm ^{-3}$. }
		\scriptsize
		\renewcommand{\arraystretch}{1.3}
		\renewcommand{\tabcolsep}{0.04cm}
		\renewcommand{\ULdepth}{6.0pt}
		\vspace{0.5cm}
		\begin{tabular}{ l|l|l|l }
			Pairing Gap & T=5, $\rho Y_e$=$10^5$ & T=30, $\rho Y_e$=10$^{11}$ & T=30, $\rho Y_e$=$10^5$ \\ \hline
			TF & 2.12$\times$10$^{-01}$ & 4.87$\times$10$^{-03}$ & 5.52 \\ 
			3TF & 3.15$\times$10$^{-01}$ & 1.13$\times$10$^{-02}$ & 9.80 \\ 
			5TF & 2.26$\times$10$^{-01}$ & 6.11$\times$10$^{-03}$ & 6.18 \\ 
		\end{tabular}
		\label{Tab8}%
	\end{table}
	\clearpage
	
	%%%%%%%%%%%%%%%%%%%%%%%%%%%%%%%%%%%%%%%%%%
	\section*{Author Contributions}{J.-U. Nabi and M. Riaz  had major contributions. A. Mehmood helped with running servers, drawing tables and figures.}
	
	\section*{Funding}{No funding received }
	
	\section*{Institutional Review}{Not Applicable}
	
	\section*{Data Availability}{Not applicable.} 
	
	\section*{Acknowledgments}
	{Authors would like to acknowledge the support of the \\ Higher Education Commission Pakistan through Project $\#$ 
		20-15394/NRPU/R\&D/HEC/2021}
	
	\section*{Conflicts of Interest}{The authors declare no conflict of interest.} 
	
	%%%%%%%%%%%%%%%%%%%%%%%%%%%%%%%%%%%%%%%%%%

	%% Only for journal Encyclopedia
	%\entrylink{The Link to this entry published on the encyclopedia platform.}
	
	\section*{Abbreviations}{
		The following abbreviations are used. \\  
		\noindent 
		\begin{tabular}{@{}l l}
			pn-QRPA & proton-neutron quasi-particle random-phase approximation\\
			GT  & Gamow-Teller \\
			TF & Traditional Formula\\
			3TF & 3 Term Formula\\
			5TF & 5 Term Formula\\
		\end{tabular}
	}
	
	%%%%%%%%%%%%%%%%%%%%%%%%%%%%%%%%%%%%%%%%%%
	%%% Optional
	%\appendixtitles{no} % Leave argument "no" if all appendix headings stay EMPTY (then no dot is printed after "Appendix A"). If the appendix sections contain a heading then change the argument to "yes".
	%\appendixstart
	%\appendix
	%\section[\appendixname~\thesection]{}
	%\subsection[\appendixname~\thesubsection]{}
	%The appendix is an optional section that can contain details and data supplemental to the main text---for example, explanations of experimental details that would disrupt the flow of the main text but nonetheless remain crucial to understanding and reproducing the research shown; figures of replicates for experiments of which representative data are shown in the main text can be added here if brief, or as Supplementary Data. Mathematical proofs of results not central to the paper can be added as an appendix.
	%
	%\begin{table}[H] 
	%\caption{This is a table caption.\label{tab5}}
	%\newcolumntype{C}{>{\centering\arraybackslash}X}
	%\begin{tabularx}{\textwidth}{CCC}
	%\toprule
	%\textbf{Title 1}	& \textbf{Title 2}	& \textbf{Title 3}\\
	%\midrule
	%Entry 1		& Data			& Data\\
	%Entry 2		& Data			& Data\\
	%\bottomrule
	%\end{tabularx}
	%\end{table}
	%
	%\section[\appendixname~\thesection]{}
	%All appendix sections must be cited in the main text. In the appendices, Figures, Tables, etc. should be labeled, starting with ``A''---e.g., Figure A1, Figure A2, etc.
	\clearpage

\end{document}